\documentclass[iop,apj,twocolappendix]{emulateapj}
\usepackage{float}
\usepackage{amsmath}
\usepackage{rotating}
\usepackage{amssymb}
\usepackage{hyperref}
\usepackage{ulem,color}

\def\gtrsim{\mathrel{\hbox{\rlap{\hbox{\lower4pt\hbox{$\sim$}}}\hbox{$>$}}}}
\def\lesssim{\mathrel{\hbox{\rlap{\hbox{\lower4pt\hbox{$\sim$}}}\hbox{$<$}}}}
\def\gtrsim{\mathrel{\hbox{\rlap{\hbox{\lower4pt\hbox{$\sim$}}}\hbox{$>$}}}}
\def\farcs{\hbox{$.\!\!^{\prime\prime}$}}
\def\farcm{\hbox{$.\!\!^{\prime}$}}

\begin{document}

\def\chan{{\sl Chandra\ }}

\title{The Mouse Pulsar Wind Nebula}

\author{Noel Klingler$^1$, Oleg Kargaltsev$^1$, George G. Pavlov$^2$, C.-Y. Ng$^3$, Paz Beniamini$^1$, Igor Volkov$^4$}
\affil{$^1$The George Washington University, Department of Physics, 725 21st St NW, Washington, DC  20052 \\ 
$^2$Pennsylvania State University, Department of Astronomy \& Astrophysics, 525 Davey Laboratory, University Park, PA 16802 \\ 
$^3$Department of Physics, the University of Hong Kong, Pokfulam, Hong Kong \\ 
$^4$University of Maryland, College of Computer, Mathematical, and Natural Sciences, College Park, MD 20742 }

\begin{abstract}
The young energetic pulsar J1747--2958 ($\tau=26$ kyr, $\dot{E}=2.5\times 10^{36}$ erg s$^{-1}$) powers the Mouse pulsar wind nebula (PWN), famous for its spectacular tail spanning 45$''$ in X-rays and 12$'$ in radio ($d\sim5$ kpc).  
We present the results of {\sl Chandra} observations of the PWN and the analysis of archival lower-frequency data.  
The {\sl Chandra} HRC image reveals a point-like source at the pulsar position, $\approx$1$'$ behind the bow shock apex of the PWN.  
The flattened appearance of the compact nebula is consistent with an  equatorial outflow deformed by the ram pressure, implying that the angle between the pulsar's spin axis and line of sight is $\sim70^\circ$ (in agreement with the radio and $\gamma$-ray pulse profiles).  
The spatially-resolved spectroscopy with {\sl Chandra} ACIS shows that the power-law (PL) spectrum steepens from $\Gamma=1.65\pm0.02$ to $3.0\pm0.1$ over the 45$'$ extent of the X-ray tail.  
However, the tail's X-ray spectrum integrated over its 45$''$ length fits a single absorbed PL with $\Gamma=2.09\pm0.03$.  
We also found the Mouse PWN in 150 MHz GMRT data, and a possible counterpart in 24 ${\rm \mu}$m {\sl Spitzer} data.  
The multiwavelength data suggest that, at low frequencies, the spectrum of the X-ray-emitting tail region can be described by a broken PL with at least one turnover between radio and X-rays.  
This is consistent with synchrotron cooling of electrons injected at the termination shock (with an SED slope of 2.2) in an equipartition magnetic field $B\sim200$ ${\rm \mu G}$ and a bulk flow speed $v\sim4000$ km s$^{-1}$.
\end{abstract}


\section{INTRODUCTION}

Pulsars are among nature's most powerful particle accelerators, producing particles with energies up to a few PeV.
As a neutron star rotates, it imparts its rotational energy into a magnetized utlrarelativistic particle wind, which encounters a termination shock (TS) downstream due to interactions with the ambient medium.
Synchrotron emission can be seen downstream of the TS from radio to $\gamma$-rays as a pulsar wind nebula (PWN). 
The observed properties of PWNe, such as their appearance, spectrum, and radiative efficiency, depend on the pulsars' parameters (e.g., the spin-down power $\dot{E}$ and the angle $\alpha$ between the magnetic and spin axes), as well as the pulsars' velocity and the properties of their environment (see reviews by Gaensler \& Slane 2006, Kargaltsev \& Pavlov 2008, Reynolds et al.\ 2017).
If a pulsar moves through the interstellar medium (ISM) at a supersonic speed, the ram pressure exerted by the ISM balances that of the pulsar wind and forms a bow shock, which redirects and channels the wind in the direction opposite to that of the pulsar's motion, forming a pulsar tail (see Bucciantini et al.\ 2005, Kargaltsev et al.\ 2017a).
Observations of fast-moving energetic pulsars in radio and X-rays have revealed such tails extending for parsecs behind the pulsars. 
In several cases, bright compact nebulae (CNe) with complex structures on small scales can be seen in the vicinity of the fast-moving pulsar. 
Studying the wind nebulae of supersonic pulsars provides information about the properties of pulsar wind (e.g., flow speeds and magnetization), particle acceleration regions, properties of the ISM and its interaction with pulsar wind, the pulsar magnetosphere geometry, and the distribution of velocities that neutron stars acquire during their progenitor supernovae.

G359.23--0.82 (``the Mouse'' PWN) is a 12$'$-long axisymmetric non-thermal nebula originally discovered in a Very Large Array (VLA) survey of the Galactic center (Yusef-Zadeh \& Bally 1987). 
Subsequent observations have revealed a bow shock structure at the head of the nebula and its synchrotron nature (Yusef-Zadeh \& Bally 1989).  
{\sl ROSAT} observations later revealed X-ray emission at the head of the Mouse (Predehl \& Kulkarni 1995). 
Follow-up observations with the Parkes radio telescope later identified the X-ray source as PSR J1747--2958: a relatively young (spin-down age $\tau_{\rm sd}=26$ kyr) and energetic ($\dot{E}=2.5\times10^{36}$ erg s$^{-1}$) pulsar powering the nebula (Camilo et al.\ 2002).
In radio, the bulbous CN and long narrow tail morphologically resemble the body and tail of a mouse, which gave rise to the PWN's nickname.
Radio polarimetry of the CN has revealed a magnetic field perpendicular to the pulsar's motion direction ahead of the pulsar, and parallel to the velocity direction in the extended tail (Yusef-Zadeh \& Gaensler 2005). 
A 36 ks {\sl Chandra} ACIS-S observation of the Mouse PWN has revealed a 45$''$ long X-ray nebula and showed that its spectrum can be described by a power-law (PL) with a photon index that increases with distance from the pulsar (Gaensler et al.\ 2004; hereafter G+04), and the 53 ks {\sl XMM-Newton} observation has shown the nebula has a spatially-averaged photon index $\Gamma\simeq1.9$ (Mori et al.\ 2005).
Pulsed $\gamma$-ray emission from PSR J1747--2958 has also been detected with the {\sl Fermi} LAT (Abdo et al.\ 2013).

The distance to the Mouse is uncertain, with somewhat different values used by different authors. 
Observations of H{\sc I} absorption showed that maximum distance of G359.23--0.82 is $\sim$5.5 kpc (Uchida et al.\ 1992). 
The Galactic electron density model of Cordes \& Lazio (2002) yields a distance of 2 kpc for the dispersion measure of the pulsar, DM = 101.5 pc cm$^{-3}$ (Camilo et al.\ 2002), while the models of both Taylor \& Cordes (1993) and Yao, Manchester, \& Wang (2017) both suggest a distance of 2.5 kpc.
G+04 noted that the hydrogen column density obtained from fits to the X-ray spectra, $N_{\rm H}\approx 2.7\times 10^{22}$ cm$^{-2}$, is much larger than the hydrogen column density implied by the the pulsar's 2 kpc DM distance, $N_{\rm H}\approx4\times10^{21}$ cm$^{-2}$.
Considering that the Galactic electron densities have large uncertainties in directions close to the Galactic center, G+04 suggested that the DM distance is unreliable.
Using the radial profile of Galactic molecular surface density from Dame (1993), G+04 found that the observed $N_{\rm H}$ indicates that the pulsar lies at a distance $\sim$ 5 kpc.
Therefore, here we also adopt the distance of 5 kpc as a reasonable estimate and will scale all distance-dependent quantities to this value.
At this distance, the proper motion of the radio CN tip measured by Hales et al.\ (2009) implies a projected transverse space velocity of $306\pm43$ km s$^{-1}$ eastward.
In Table 1 we the list relevant pulsar parameters.

In this paper we present the study of a series of deeper X-ray observations of the Mouse PWN taken with the {\sl Chandra X-ray Observatory (CXO)}.
In Section 2 we describe the observations and data reduction techniques used.
In Section 3 we present X-ray and radio images of the PWN, spectral maps, and in-depth spectral fit results.
In Section 4 we discuss the implications of our findings and examine the multiwavelength properties of the Mouse, and in Section 5 we present our conclusions.

\section{OBSERVATIONS, DATA REDUCTION, AND DATA ANALYSIS}

Four $\approx$30 ks observations of the Mouse PWN were carried out over the course of 16 months, from March 2013 to July 2014.
The data were taken with the {\sl CXO's} ACIS-I instrument operating in Very Faint timed exposure mode (3.2 s time resolution).
In our analysis we also used the archival 36 ks ACIS-S observation (ObsID: 2834; PI: Gaensler), which was taken in Faint timed exposure mode, and the archival 58 ks HRC-I observation (ObsID: 9106; PI: Ng). 
The details of these observations are summarized in Table \ref{tbl-obs}.
Only the ACIS observations are used for spectral analysis, as the HRC observation has (almost) no spectral information. 
It is worth noting that there is one additional archival {\sl Chandra} observation which contains the Mouse within its field of view: ObsID 14596 (ACIS-S); however this observation is not used in our analysis as the Mouse was too far from the optical axis ($\sim9'$) to allow any meaningful spectral or morphological analysis of the PWN (due to blurring caused by the much broader off-axis PSF, which would preclude spatially-averaged spectroscopy of the PWN as it would be contaminated from the pulsar emission).

\begin{figure*}
\epsscale{1.15}
\plotone{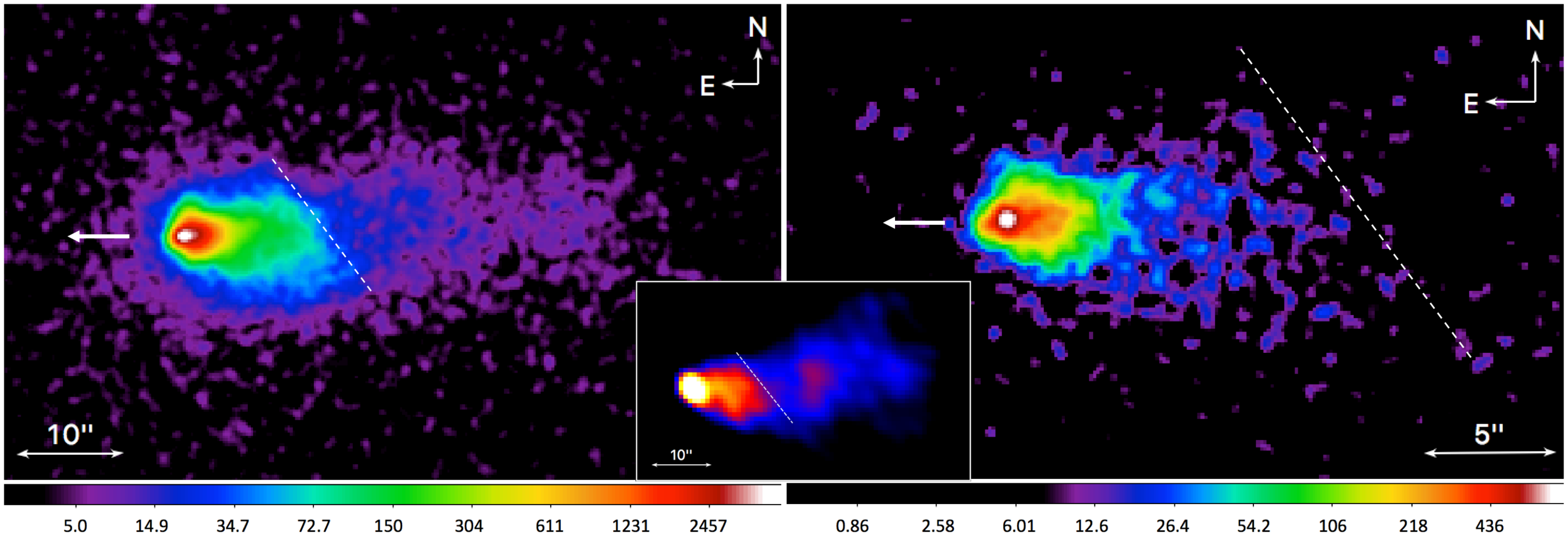}
\caption{{\sl Left}:  Merged image from five ACIS observations (0.5--8 keV; the total exposure is 154 ks) with a pixel size of $0\farcs25$ (a binning factor of 0.5 applied to the native $0\farcs492$ pixels) and smoothed with an $r=0\farcs75$ (3 pixel) Gaussian kernel.  {\sl Right}:  Zoomed-in HRC-I image (58 ks; pixel size $0\farcs123$) smoothed with an $r=0\farcs37$ (3 pixel) Gaussian kernel.  The white arrows show the direction of the pulsar's proper motion (eastward) measured by Hales et al.\ (2009).  The dashed white line marks the sharp drop in surface brightness seen in the ACIS image.  {\sl Inset}: VLA radio image (1.5 GHz) of the Mouse CN which shows   a brightness drop at approximately the same inclination as the one seen in the ACIS image.}
\label{fig-images}
\end{figure*}

We used the \chan Interactive Analysis of Observations (CIAO) software version 4.8 and {\sl CXO} Calibration Database (CALDB) version 4.7.0 for all data processing.
All observations were reprocessed using {\tt chandra\_repro} to ensure that the latest calibrations are applied.  
For the ACIS observations, we applied boresight corrections to account for small systematic offsets between the observations. 
The corrections were determined by binning the ACIS images by a factor of 0.25 (of the native ACIS pixels of size $0\farcs492$), aligning the pulsar position\footnote{The centroid position was calculated using counts within an $r=0\farcs55$ circle centered on the brightest pixel.} to that in the longest ACIS observation (ObsID 2834), and applying CIAO's {\tt wcs\_update} to modify the WCS information in the file headers. 
To improve the absolute {\sl Chandra} astrometry, we used {\tt wcs\_match} to minimize the positional offsets between the X-ray centroid positions of nearby field point sources (obtained by running {\tt wavdetect} on the longest observation, ObsID 2834) and the positions of their 2MASS counterparts (within 1$''$ of the X-ray sources).
The resulting aligned astrometry-corrected images were used to produce the merged ACIS image shown in Figure \ref{fig-images} together with the HRC-I image. 
We created exposure maps for each observation and produced a merged exposure-map-corrected image with the {\tt merge\_obs} routine using the default effective energy of 2.3 keV.  
Spectra were extracted using the CIAO {\tt specextract} tool and were fitted with absorbed power-law (PL) models in XSPEC with solar abundances assumed (we used the {\tt tbabs} model, which uses the absorption cross sections from Wilms et al.\ 2000).
For the ACIS observations, photon energies were restricted to the 0.5--8 keV range for all images and spectral analyses.  
All uncertainties listed below are at the 1$\sigma$ confidence level unless specified otherwise.

To create the adaptively-binned spectral map, we use the weighted Voronoi tessellation (WVT) algorithm of Diehl \& Statler (2006), which is a generalization of the Voronoi binning algorithm by Cappellari \& Copin (2003).  
WVT is a method of adaptively binning two-dimensional data so that each bin (spatial region) meets a specified signal-to-noise ratio (S/N) requirement (i.e., the spatial resolution is maximized while a constant S/N is maintained across the spatial bins).
Diehl \& Statler (2006) developed a method to create adaptively-binned temperature maps of supernova remnants and galaxy clusters\footnote{The source code can be found at http://www.phy.ohiou.edu/\~diehl/WVT/}.
We adapted their code for the creation of spectral maps of PWNe from non-adjacent (in time) ACIS observations (see Kargaltsev et al.\ 2017b). 
We applied all required observation-specific calibration corrections (as mentioned above) and performed simultaneous spectral fits for the same spatial region from different observations; this allowed us to use individual, observation-specific responses for each region.
To create the WVT bins for the spectral map, we use the merged ACIS image binned by a factor of 0.5 from the native ACIS pixels (to a new pixel size of $0\farcs25$), and require a S/N of 30 for each WVT bin (spatial region).

Radio images were obtained from the NRAO VLA Archive Survey (NVAS; Crossley et al.\ 2007)\footnote{http://archive.nrao.edu/nvas/}.  
The 150 MHz flux density was measured from the pipeline-processed image from the TIFR (Tata Institute of Fundamental Research) GMRT (Giant Metrewave Radio Telescope) Sky Survey (TGSS; Intema et al.\ 2017)\footnote{http://tgssadr.strw.leidenuniv.nl/doku.php}, and the infrared (IR) flux was measured from a pipeline-processed image from the {\sl Spitzer} MIPSGAL 24 ${\rm \mu}$m Galactic Plane Survey (Gutermuth \& Heyer 2015).
We used the Python package {\tt naima} (Zabalza 2015) to plot the multiwavelength spectra.

\begin{deluxetable}{lc}
\tablecolumns{9}
\tablecaption{Observed and Derived Pulsar Parameters \label{tbl-parameters}}
\tablewidth{0pt}
\tablehead{\colhead{Parameter} & \colhead{Value} }
\startdata
R.A. (J2000.0; radio position), $\alpha$ & 17 47 15.882 (8)  \\  
Decl. (J2000.0; radio position), $\delta$ & --29 58 01.0 (7)  \\
Epoch of position (MJD) & 52613  \\
Galactic longitude, $l$ (deg) & 359.31  \\ 
Galactic latitude, $b$ (deg) & --0.84  \\ 
Spin period, $P$ (ms) & 98.8  \\
Spin period derivative, $\dot{P}$ ($10^{-14}$ s s$^{-1}$) & $6.132$ \\
Dispersion measure, DM (pc cm$^{-3}$) & 101.5  \\ 
Adopted distance, $d$ (kpc) & 5  \\ 
Velocity, $v_\perp$ (km s$^{-1}$) & $306\pm43$ \\ 
Spin-down power, $\dot{E}$ (erg s$^{-1}$) & $2.5\times10^{36}$  \\ 
Spin-down age, $\tau_{\rm sd} = P/(2\dot{P})$ (kyr) & 25.5 \\
Surface magnetic field, $B_{\rm surf}$ (G) &  $2.5\times10^{12}$
\enddata
\tablenotetext{}{Parameters are from the ATNF catalog (Manchester et al.\ 2005, Camilo et al.\ 2002, G+04, and Hales et al.\ 2009).  The values in parentheses are the last digit errors for the pulsar position.}
\end{deluxetable}

\begin{deluxetable}{ccccc}
\tablecolumns{4}
\tablecaption{\chan observations used in our analysis
\label{tbl-obs}}
\tablewidth{0pt}
\tablehead{\colhead{ObsId} & \colhead{Instrument} & \colhead{Exposure} & \colhead{Date} & \colhead{$\theta$} \\ \colhead{} & \colhead{} & \colhead{(ks)} & \colhead{} & \colhead{(arcmin)}  }
\startdata
2834 & ACIS-S & 36.3 & 2002 Nov 19 & 1.23 \\
9106 & HRC-I & 57.8 & 2008 Feb 07 & 0.26  \\
14519 & ACIS-I & 29.7 & 2013 Mar 28 & 0.56  \\
14520 & ACIS-I & 30.6 & 2013 Jul 25 & 0.58  \\
14521 & ACIS-I & 27.8 & 2014 Mar 16 & 0.66  \\
14522 & ACIS-I & 29.7 & 2014 Jul 25 & 0.68 
\enddata
\tablenotetext{}{$\theta$ is the angular distance between the pulsar and the optical axis of the telescope.}
\end{deluxetable}

\section{RESULTS}

\subsection{Spatial Morphology}
The 118 ks \chan data from the new observations offer a much deeper view of the Mouse PWN compared to the 36 ks observation described by G+04.
The HRC and ACIS images, at their highest resolution, both reveal an interesting morphology within a few arcseconds of the pulsar.  
The head of the CN (i.e., the brighter area upstream (east) of the dashed line in Figure 1) displays a cometary shape with a ``filled'' morphology, but the front edge of the bright CN does not follow a simple parabolic shape (see the green-colored areas in Figure \ref{fig-images}). 
The HRC-I image best shows the excess emission ahead of the pulsar (the emission east of the white-colored area coincident with the pulsar) accompanied by the rapid widening of the CN that is particularly pronounced north of the pulsar and seen up to about $4''$ from it. 
The CN continues to widen up to about $10''$ from the pulsar (in the direction opposite to that of the pulsar motion). 
This behavior abruptly changes beyond this distance, where the nebula becomes fainter and narrower. 
The transition between the two regimes happens along the boundary which is shown by the inclined dashed line in Figure \ref{fig-images} (left panel) which has a position angle of $37^\circ$ degrees East of North.

\begin{figure}
\epsscale{1.1}
\plotone{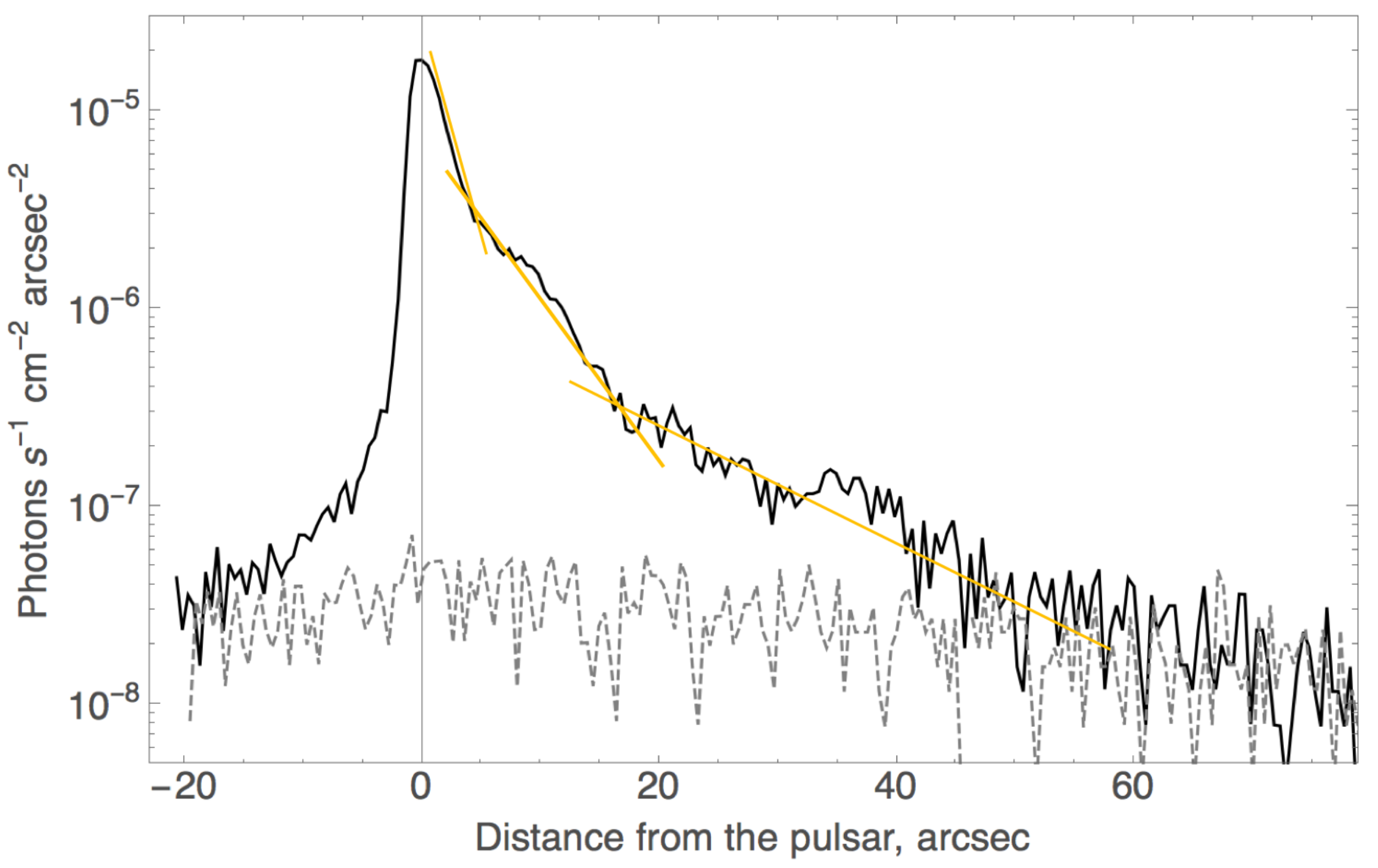}
\caption{Linear brightness profile of the Mouse PWN obtained from the exposure-corrected ACIS image (in 0.5--8 keV), using a rectangular extraction region with a height of 12$''$ in the North-South direction (vertically centered on the pulsar) and a length of 80$''$ in the East-West direction (West being represented by positive distance values, and zero corresponding to the pulsar position).  The dashed gray curve shows the background surface brightness, obtained using the same extraction region placed outside the PWN ($20''$ to the North).  
The orange lines are plotted over the brightness profile to highlight the three segments with different dependences of brightness on distance from the pulsar.}
\label{fig-brightness-profile}
\end{figure}

In Figure \ref{fig-brightness-profile} we present the brightness profile of the PWN, obtained from the exposure-corrected merged ACIS image (using a $12''\times100''$ region with the longer dimension oriented along the east-west direction and centered on the pulsar in the transverse direction).
Three distinct regimes can be identified in the brightness profile based on the change of the profile slope (shown by the straight orange lines in Figure \ref{fig-brightness-profile}).  
Note that the vertical scale is logarithmic and, therefore, a linear trend in the plot corresponds to an exponential (or nearly exponential) decay. 
The X-ray tail brightness fades to background levels approximately 50$''$ from the pulsar. 

\begin{figure*}
\epsscale{1.15}
\plotone{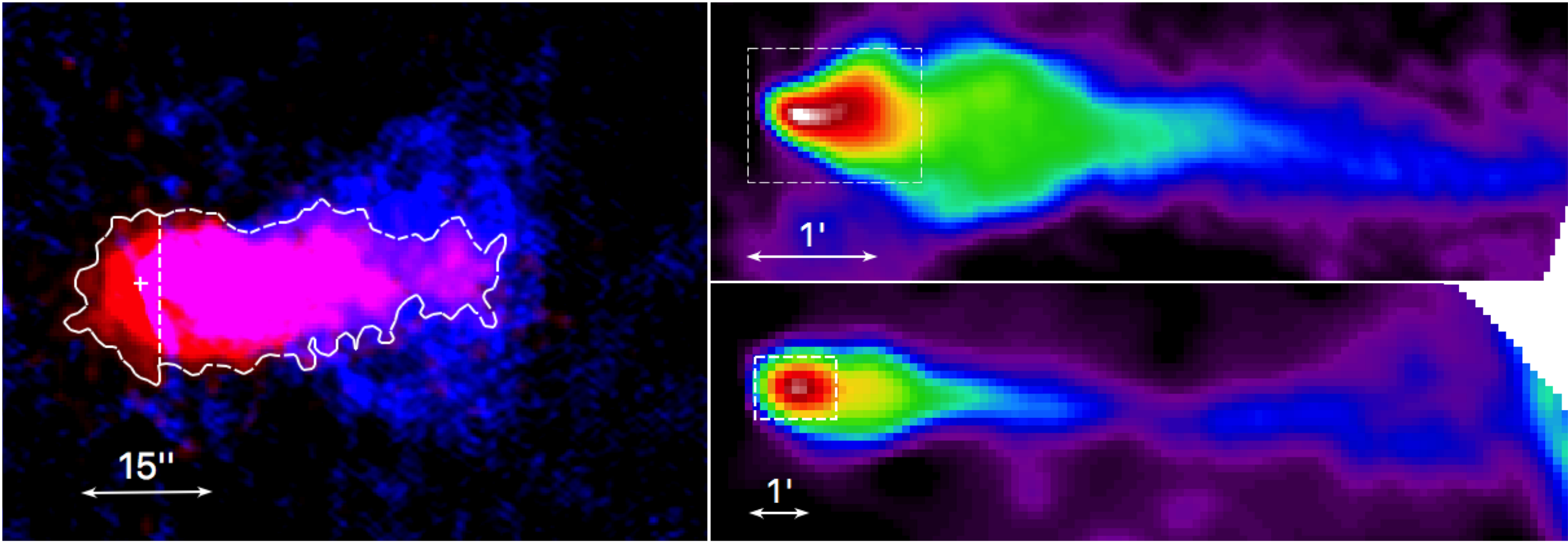}
\caption{{\sl Left}:  Composite X-ray (red; binned by a factor of 0.5, and smoothed with an $r=0\farcs74$ (3-pixel) Gaussian kernel) and radio (blue; VLA, 4.77 GHz, $1\farcs17\times1\farcs00$ beam) image of the Mouse PWN.  The radio pulsar position is marked by the cross.  The white  contour encloses the X-ray nebula with the western (dashed) segment used to extract the entire tail spectrum (Section 3.3) and  multiwavelength spectrum (Section 4).  {\sl Right}: VLA radio images (top: 4.89 GHz, $12\farcs4\times9\farcs8$ beam; bottom: 1.49 GHz, $34\farcs6\times30\farcs5$ beam) showing the extended tail of the Mouse.  The field of view of the image on the left is shown  by the dotted white box in the right panels.  The radio images were obtained from the NRAO VLA Archive Survey (Crossley et al.\ 2007).}
\label{fig-radio}
\end{figure*}

\subsection{Spatially-Resolved Spectra}

The sub-arcsecond resolution of \chan allows us to perform spatially-resolved spectroscopy despite the small angular size of the PWN. 
We first explored the pulsar vicinity to see if there is any spectral evidence of the pulsar being detected and then characterized the spectral changes in the extended PWN. 
In all cases we simultaneously fit the spectra (0.5--8 keV) from all 5 ACIS observations of the Mouse PWN (since no significant spectral nor morphological changes are seen in the PWN between the observations).
We obtained the background spectrum using a rectangular 138$''\times39''$ region placed well outside the boundaries of both the X-ray and radio tails (centered at R.A.=17$^{\rm h}$47$'07\farcs5$, Decl.=--$29^{\circ}56'43\farcs2$, and excluding two faint field point sources), and subtracted it from the source spectra in all spectral fits.

\subsubsection{Spectrum in the Pulsar's Vicinity}

To probe the pulsar's spectrum, we used a small extraction region of radius $r=0\farcs74$ (1.5 native ACIS pixels, corresponding to approximately a 90\% encircled counts fraction\footnote{see http://cxc.harvard.edu/ciao/PSFs/psf\_central.html}), centered at the brightest pixel when the merged image is binned by a factor of 0.5.
The small aperture is intended to minimize the contamination from the nebula. 
The extracted pulsar spectrum was then fitted with an absorbed PL model which produced a good fit: $\chi_\nu^2=1.09$ for $\nu=99$ degrees of freedom (dof).
The best-fit $N_{\rm H}=(2.61\pm0.18)\times10^{22}$ cm$^{-2}$ is consistent with the value previously reported by G+04 ($N_{\rm H,22}=2.7\pm0.1$ cm$^{-2}$), and with the value found from fitting the entire X-ray tail ($N_{\rm H,22}=2.8\pm0.07$ cm$^{-2}$, see Section 3.2.2), so we fixed $N_{\rm H,22}=2.7$ for all subsequent fits.
The fit of the pulsar's spectrum yields photon index $\Gamma=1.55\pm0.04$ and an absorbed X-ray flux of $(1.04\pm0.01)\times10^{-12}$ erg s$^{-1}$ cm$^{-2}$.  
Using {\sl Chandra} PIMMS and the above fit parameters, we estimate that the pile-up fraction does not exceed 7\%. 
Attempts to fit the pulsar spectrum with a PL+blackbody (PL+BB) model with $\Gamma$ frozen at 1.55 and a BB normalization corresponding to an $R=10$ km NS at 5 kpc yields $kT=131\pm27$ eV, but the fit quality is virtually the same as for the PL-only fit ($\chi^2_\nu=1.10$ for $\nu=98$), and the thermal component is therefore not statistically required by the data. 
We found an upper limit on the neutron star surface temperature by gradually increasing the temperature of the BB component (and re-fitting the PL component at each new temperature) until the model became inconsistent with the data at the 3$\sigma$ confidence level.
This yields an upper limit $kT=166$ eV ($T=1.93$ MK).

\subsubsection{PWN}

\begin{figure}
\epsscale{1.15}
\plotone{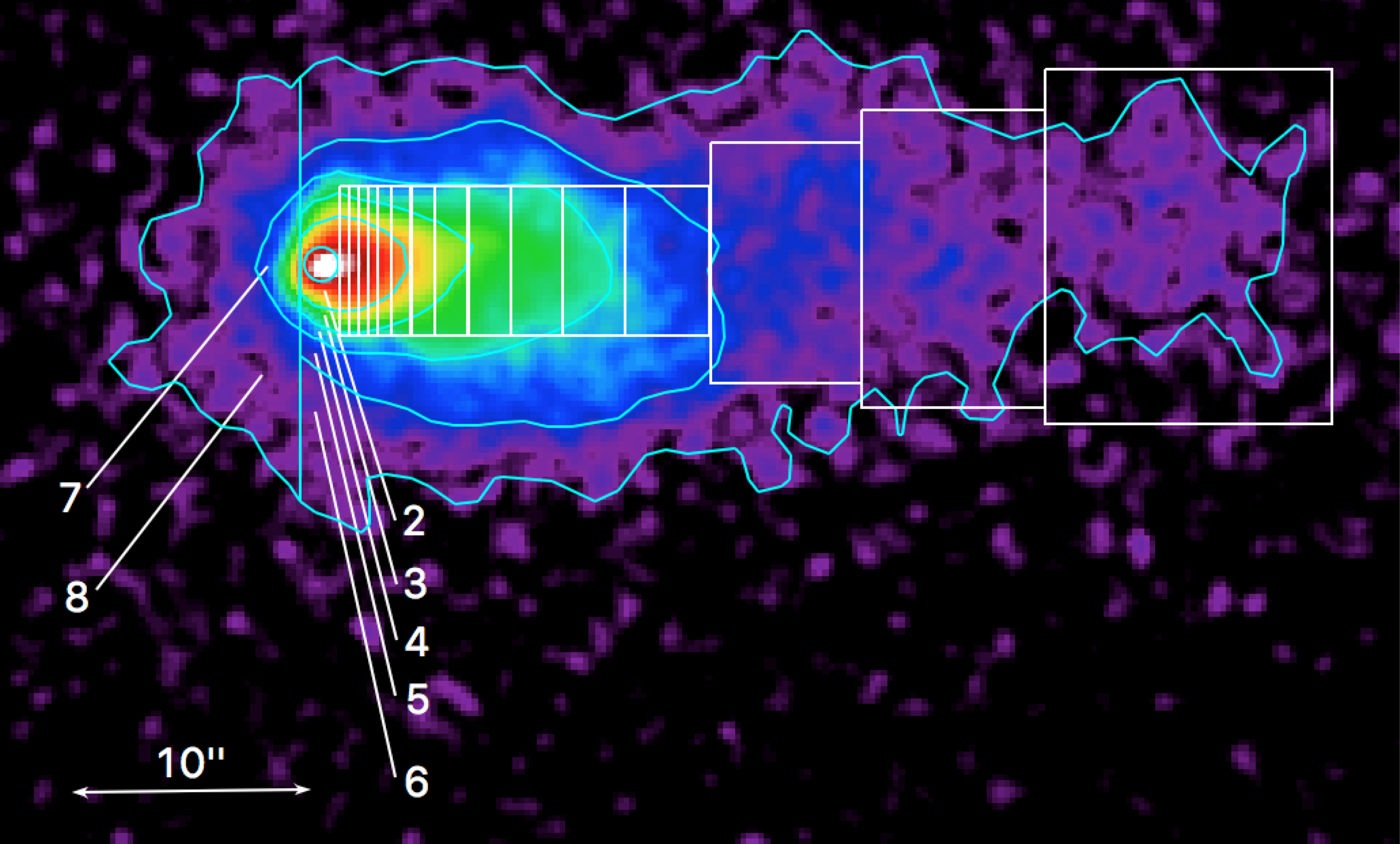}
\caption{Merged ACIS counts image (the same as in the left panel of Figure 1) showing the regions used for spectral analysis (see Section 3.3 for details) -- the boxes (white) and the numbered contour regions (cyan): Contour 2 (the red-colored area in the vicinity of the pulsar), Contour 3 (yellow area), Contour 4 (green area), Contour 5 (light blue and blue area), and Contour 6 (blue and purple area); Contours 7 (predominantly light blue) and 8 (purple) cover the area ahead of the pulsar.} 
\label{fig-regions}
\end{figure}

We performed spatially-resolved spectroscopy for the rest of the PWN in several ways.
We first followed the approach of G+04 and extracted spectra from regions enclosed within the brightness contours (we will call them ``contour regions'' below; see Figure \ref{fig-regions}). 
The spectra fitted with an absorbed PL model (see Table \ref{spatially-resolved-spectra}) exhibit substantial changes in the photon index, $\Delta\Gamma\approx1$, between the inner and the outer contour regions (regions 2 and 6 in Figure \ref{fig-regions}).  
The two outermost contour regions (7 and 8 in Figure \ref{fig-regions}) were divided into two parts along the North-South line placed just ahead of the pulsar to determine if the spectrum changes ahead of the pulsar.  
The inner head (within region 7) exhibits a significantly harder spectrum, $\Gamma=1.54\pm0.16$, than that of the outer head (within region 8), which has $\Gamma=2.61\pm0.15$.

\begin{deluxetable*}{lccccccccc}
\tablecolumns{4}
\tablecaption{Spectral Fits for Different Regions
\label{spatially-resolved-spectra}}
\tablewidth{0pt}
\tablehead{\colhead{Contour Region} & \colhead{Area} & \colhead{Net Counts,} & \colhead{Counts per bin} & \colhead{$\Gamma$} & \colhead{$\mathcal{N}_{-4}$} & \colhead{$\chi_\nu^2$ ($\nu$)}  & \colhead{$F_{X,-12}$} & \colhead{$L_{X,33}$} \\ 
\colhead{} & \colhead{(arcsec$^2$)} & \colhead{0.5--8 keV} & \colhead{} & \colhead{} & & & \colhead{}}
\startdata
Pulsar & 1.7 & $7031\pm84$ & 60 & $1.55\pm0.04$ & $3.01\pm0.14$ & 1.09 (99) & $1.95\pm0.03$  &  $5.83\pm0.09$  \\ 
2 & 12.4 & $16803\pm130$ & 120 & $1.65\pm0.02$ & $8.10\pm0.23$ & 1.12 (120) & $4.77\pm0.05$ & $14.3\pm0.3$ \\ 
3 & 17.4 & $5223\pm73$ & 60 & $1.86\pm0.04$ & $3.05\pm0.17$ & 1.75 (68) & $1.50\pm0.04$ & $4.48\pm0.12$ \\ 
4 & 52.5 & $5390\pm74$ & 60 & $2.19\pm0.04$ & $4.78\pm0.23$ & 1.14 (74) & $1.88\pm0.04$ & $5.62\pm0.12$ \\ 
5 & 100.5 & $3380\pm59$ & 50 & $2.25\pm0.05$ & $3.10\pm0.20$ & 1.11 (52) & $1.18\pm0.05$ & $3.53\pm0.15$ \\ 
6 & 159.1 & $3145\pm57$ & 30 & $2.69\pm0.05$ & $4.44\pm0.25$ & 1.10 (91) & $1.42\pm0.06$ & $4.25\pm0.18$ \\ 
7 (inner head) & 6.7 & $582\pm24$ & 15 & $1.54\pm0.16$ & $0.26\pm0.05$ & 1.58 (30) & $0.17\pm0.01$ & $0.51\pm0.03$ \\ 
8 (outer head) & 72.6 & $544\pm25$ & 12 & $2.61\pm0.15$ & $0.69\pm0.12$ & 1.24 (35) & $0.23\pm0.03$ & $0.78\pm0.18$ \\ 
Entire X-ray tail & 530.9 & $22490\pm150$ & 75 & $2.09\pm0.02$ & $16.2\pm0.03$ & 0.98 (242) & $6.78\pm0.07$ & \ \ \ $20.3\pm0.2$ 
\tablenotetext{}{Best-fit spectral parameters for an absorbed PL model for the contour regions shown in Figure \ref{fig-regions}.  $N_{\rm H}$ was fixed at $2.7\times 10^{22}$ cm$^{-2}$.  The unabsorbed flux $F_{X,-12}$ and luminosity $L_{X,33}$ are given in units of $10^{-12}$ erg cm$^{-2}$ s$^{-1}$ and $10^{33}$ erg s$^{-1}$, respectively, in the 0.5--8 keV range.  The luminosity is calculated for a distance of 5 kpc.
$\mathcal{N}_{-4}$ is the PL model normalization in units of $10^{-4}$ photons s$^{-1}$ cm$^{-2}$ keV$^{-1}$ at 1 keV, and counts per bin refers to the number of counts in each energy bin.  Note that the entire X-ray tail region (Figure 3, left panel) is not the same as regions 2--6 combined.}
\enddata
\end{deluxetable*}

\begin{figure}
\epsscale{1.15}
\plotone{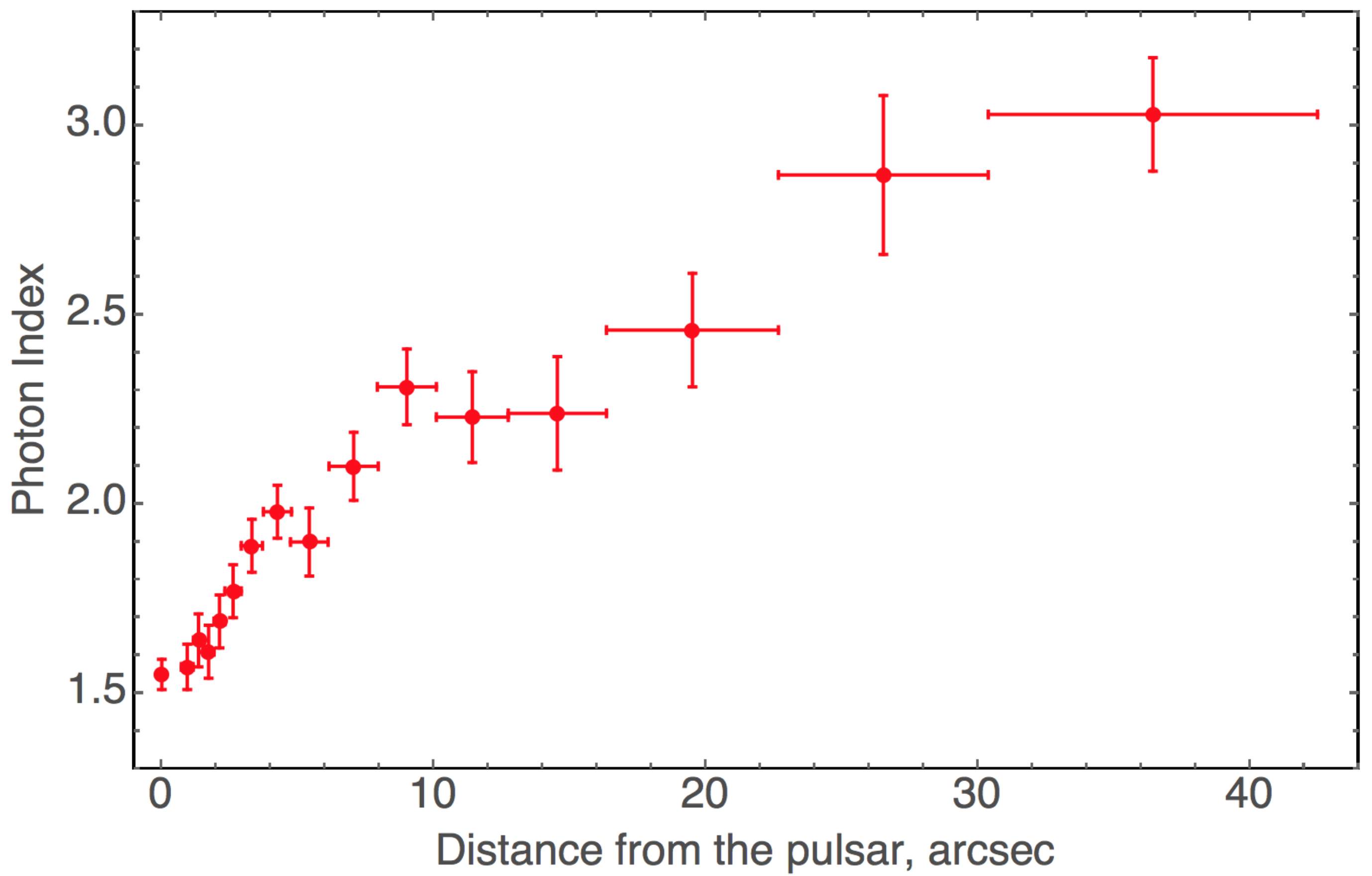}
\caption{Photon index $\Gamma$ as a function of distance from the pulsar.  The points show the spectra extracted from the rectangular regions shown in Figure \ref{fig-regions}, with the left-most point corresponding to the pulsar.  The horizontal error bars show the spatial extent (the width of each rectangle in the direction along the tail) of the regions from which the spectra were extracted, and the vertical error bars show 1$\sigma$ uncertainties in the spectral slope $\Gamma$.}
\label{fig-cooling}
\end{figure}

To further investigate spectral softening with distance from the pulsar along the X-ray tail, we extracted spectra from 15 rectangular regions shown in Figure \ref{fig-regions}. 
Since the tail's brightness decreases with distance, we used larger regions at the end of the tail to collect enough counts.
We found that $\Gamma$ increases nearly monotonically with distance, as seen in Figure \ref{fig-cooling}. 
The photon index $\Gamma$ in the outermost rectangular region is larger than that of the innermost one by $\Delta\Gamma\approx1.4$.
This variation is noticeably larger than that found for the contour regions (see Table 3), perhaps because the outermost regions include areas of harder spectra closer to the pulsar.

Fitting the spectrum of the entire X-ray tail (the dashed contour in Figure \ref{fig-radio}) produces a statistically acceptable fit with $\Gamma = 2.09 \pm 0.02$ and $\chi_\nu^2=0.98$ for $\nu=242$.  
This underscores the danger of inferring the slope $p$ of the spectral energy distribution (SED) of the particles injected at the termination shock from the photon index $\Gamma=(p+1)/2$ (for synchrotron radiation) obtained from a fit to the spectrum extracted from a large region of a PWN (a common choice for fainter PWNe necessitated by low count statistics; $p=2\Gamma-1$).
Leaving $N_{\rm H}$ as a free parameter in the above fit yields $N_{\rm H}=(2.80\pm0.07)\times10^{22}$ cm$^{-2}$ (consistent with the value obtained above from fitting the spectrum from the pulsar vicinity), and $\Gamma=2.14\pm0.04$ with $\chi_\nu^2=0.97$ for $\nu=241$ d.o.f.

\begin{figure*}
\epsscale{1}
\plotone{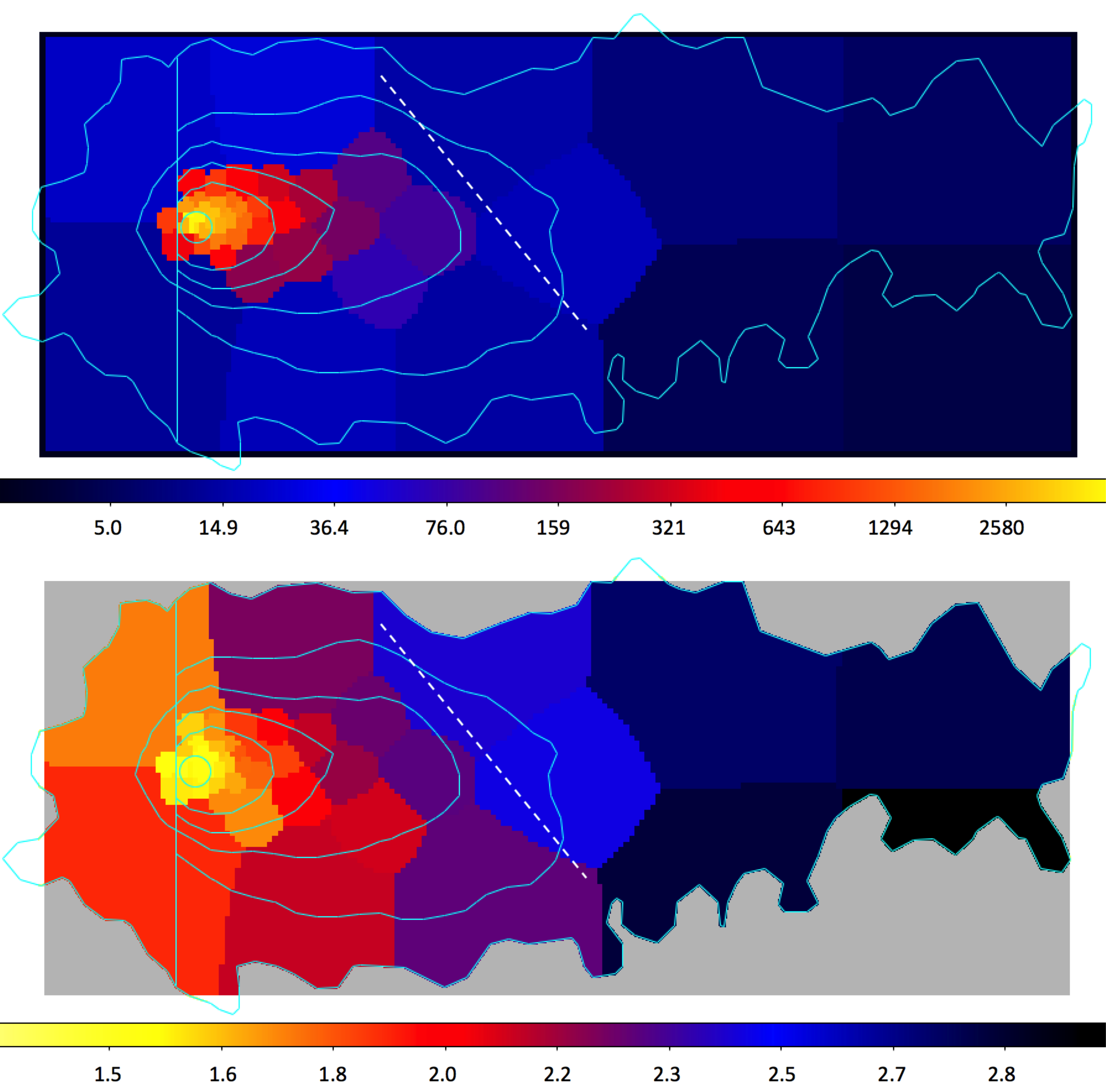}
\caption{Top: Adaptively-binned merged ACIS image of the Mouse PWN (see Section 2).  
The color bar shows brightness in units of counts arcsec$^{-2}$.  Bottom: Adaptively-binned spectral map of the Mouse PWN.  The color bar shows the photon index.  
The cyan brightness contours (the same as in Figure \ref{fig-regions}) are overlaid, the dashed white line is the same from Figure 1, and the dashed lineand the areas outside of them are grayed out for illustrative purposes.}
\label{fig-spmap}
\end{figure*}

Finally, in order to investigate the spectral variations within the Mouse PWN on even smaller spatial scales, we produced adaptively-binned spectral maps. 
The weighted Voronoi tesselation (WVT) approach varies the bin sizes which allows us to achieve the maximum spatial resolution for the specified S/N=30.
We first binned the merged ACIS image into regions such that they have a minimum S/N of 30 (Figure \ref{fig-spmap}, top panel)\footnote{Some regions will have S/N $>$ 30 since the addition of bright pixels to a region can place the S/N above the minimum} and then fit the spectra (from the five individual observations) for each region to create an adaptively-binned spectral map (Figure \ref{fig-spmap}, bottom panel). 
The spectral map clearly shows the spectral softening in front of and behind the pulsar (see above) as well as some hints of lateral softening.

\subsection{HRC Timing}

To search for X-ray pulsations, we performed the $Z_n^2$ ($n=1$,2) periodicity search (Buccheri et al.\ 1983) in the archival HRC data (ObsID 9106). 
Prior to searching, we corrected the event arrival times from the pipeline-produced Level 2 event list to the solar system barycenter using the CIAO task {\tt axbary}.  
We restricted the search to $N=350$ events located within $r=0\farcs4$ centered on the brightest pixel of the HRC image of the Mouse PWN. 
The choice of the region size was dictated by the distance up to which the PWN core maintains an approximately round shape consistent with that of a point source.  

The observation time span, $T_{\rm span}=57,760$ s, provides $\sim 1/T_{\rm span}=17.3$~$\mu$Hz resolution.  
There is a known wiring error with the HRC-I detector which causes the photon arrival times to be associated with that of the following event, which may or may not have been telemetered, thus degrading the timing accuracy to roughly the mean time between events\footnote{see http://cxc.harvard.edu/cdo/hrc/timing.html for details.}.
Since there were 3,627,988 total counts recorded by the entire detector during the 57.76 ks observation, the source photon arrival times are expected to be off by 15.92 ms on average. 
This should further degrade the resolution by broadening $Z_n^2$ peaks.

We found the maximum values $Z_1^2=3.0$ and $Z_2^2=6.8$ in the $\pm18$~$\mu$Hz range around the expected $\nu=10.1190144683(5)$ Hz (at MJD 54503.9) based on the {\sl Fermi} LAT ephemeris\footnote{https://fermi.gsfc.nasa.gov/ssc/data/access/lat/ephems/ lat\_psrcat/radio/ephem\_J1747-2958\_gbt.par}.    
We chose to search over a frequency range that is wider than the uncertainty of the predicted frequency because the pulsar is known to exhibit glitches according to the the {\sl Fermi} LAT ephemeris.
We note that the epoch of the HRC observation is outside the validity ranges of both the Fermi LAT (MJD 54634.203--54941.497) and ATNF (MJD 52306--52918) ephemerides. 
We picked the Fermi LAT ephemeris for the frequency prediction because it is closest to the HRC observation date. 
Using the more distant ATNF ephemeris results in the predicted frequency of $10.119000323(2)$ Hz which is 14.1 $\mu$Hz away from the  Fermi LAT ephemeris prediction. 

We estimated the upper limit on the observed pulsed fraction $p_{\rm obs}$ at 99.73\% confidence (3$\sigma$) by solving Equation (8) of Fierro et al.\ (1995) and found $p_{\rm obs}<34$\%.
However, the intrinsic pulsed fraction limit could be higher due to the presence of high background contamination from the surrounding nebula.
From the surface brightness of the nebula in the immediate vicinity of the pulsar, we estimate that out of the $N_{\rm tot}=350$ counts extracted from the region centered on the pulsar, roughly half of the counts ($N_{\rm bkg}=170$) come from the nebula.
Thus, the upper limit on the intrinsic pulsed fraction could be as high as $p_{\rm int}=p_{\rm obs}(1+ N_{\rm bkg}/N_{\rm tot})=51\%$.

\section{Discussion}

The new deeper \chan observations of the Mouse PWN have allowed us to better characterize its spatial and spectral morphology.  
Together with the multiwavelength data described below, they provide additional constraints on the PWN parameters and on the pulsar magnetosphere geometry.

\subsection{Linking Compact PWN Morphology with Magnetosphere Geometry}

Pulsars which share similar pulse profiles in both radio and $\gamma$-rays are likely to have similar viewing angles $\zeta$ (between the line of sight and pulsar spin axis) and similar magnetic inclination angles $\alpha$ (between the spin and magnetic dipole axes), as current models of magnetospheric emission suggest that the observed pulse profile shapes are primarily determined by these two angles (see, e.g., Muslimov \& Harding 2004). 
Radio and $\gamma$-ray pulse profiles have been used to infer $\alpha$ and $\zeta$ for different magnetosphere models (e.g., Romani \& Watters 2010; Pierbattista et al.\ 2016).

One might expect that PWNe powered by pulsars with similar $\zeta$ and $\alpha$ will resemble each other; i.e., there will be a correlation between the PWN morphologies and the pulse profiles of pulsars powering these PWNe. 
However, the differing magnitudes and directions of pulsar velocities can have a significant impact on the PWN morphologies resulting in different PWN appearances even when $\zeta$ and $\alpha$ are similar. 
This is particularly important for supersonically-moving pulsars (see Kargaltsev et al.\ 2017a).
For example, PSRs J1509--5058 and J1709--4429 (B1706--44) have remarkably similar pulse profiles (see Figure \ref{fig-lightcurve}; middle and bottom panels). 
The similarities include the separations between the radio and $\gamma$-ray peaks, the double-peaked morphologies, as well as the widths of the $\gamma$-ray pulses and the relative strength and the separation between the two $\gamma$-ray peaks. 
Although the X-ray PWNe of these two pulsars appear to be quite different at first glance, the similarities can be identified after accounting for differences in the pulsar velocities. 
According to Romani et al.\ (2005), the transverse velocity of PSR J1709--4429 (B1706--44) is $\lesssim 40$ km s$^{-1}$ while for PSR J1509--5058 the velocity is substantially larger, $160$--$640$ km s$^{-1}$ (Klingler et al.\ 2016a).  
The larger velocity can explain the more curved jets and much more collimated extended tail of PSR J1509--5058 (see Klingler et al.\ 2016a). 
Yet, in both cases one can see two jets of roughly-equivalent brightness (implying a large $\zeta$) and relatively dim tori compared to the jets (suggesting that $\sin\alpha$ is sufficiently small).
In fact, for PSR J1709--4429, $\zeta=53\pm0.4^\circ$ was measured by Romani et al.\ (2005) by fitting a Doppler-boosted torus model to the {\sl Chandra} ACIS image of the compact PWN.

The radio and $\gamma$-ray light curves of the Mouse pulsar (J1747--2958) also share similarities with those of PSRs J1509--5850 and J1709--4429 (see Figure \ref{fig-lightcurve}).  
Here, the only differences are that the Mouse's $\gamma$-ray pulse profile is slightly wider (with a deeper trough) and more asymmetric compared to those of J1509 and J1709 (see Figure \ref{fig-lightcurve}).  
In radio, all three pulsars display single peaks with very similar phase separation from the $\gamma$-ray pulses.   
According to the Outer Gap magnetospheric emission model (e.g., Romani \& Watters 2010), this implies a fairly large $\zeta$ and a similarly large $\alpha$ (to ensure that both $\gamma$-ray and radio pulsations can be seen). 
The stronger asymmetry and a deeper minimum between the pulses can be attributed to a large $\alpha\sim70^{\circ}-90^{\circ}$, which is also a plausible range for $\zeta$. 
These angles are likely to be somewhat smaller for J1747 than those of PSRs J1509 and J1709.  
Comparison of the J1747 $\gamma$-ray and radio pulse profiles with simulated pulse profiles using the Two Pole Caustic (TPC) and Outer Gap (OG) $\gamma$-ray emission models (see Figure 2 of Watters et al.\ 2009) suggest the plausible ranges of angles 
$\zeta_{\rm OG}\sim$ 58$^{\circ}-78^\circ$, $\zeta_{\rm TPC}\sim 40^{\circ}-71^\circ$, $\alpha_{\rm OG}\sim 50^{\circ}-67^\circ$, and $\alpha_{\rm TPC}\sim 39^{\circ}-70^\circ$.

These considerations help to interpret the appearance of the CN discussed in the next section.

\subsection{PWN Core and Pulsar}

\begin{figure}
\epsscale{1.2}
\plotone{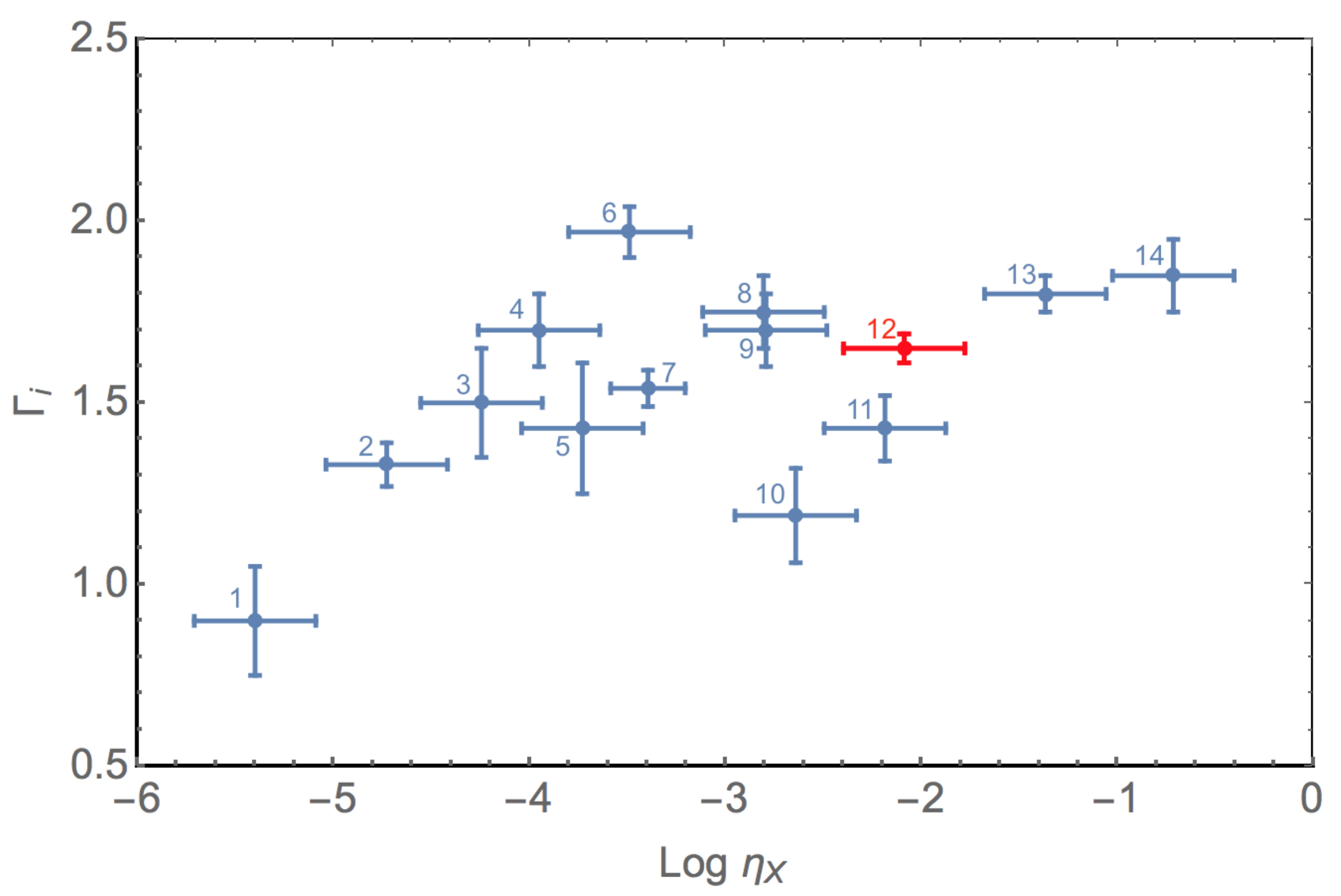}
\caption{Slopes $\Gamma_i$ of the X-ray spectra emitted by particles injected at the termination shock vs PWN X-ray (0.5--8 keV) efficiency $\eta_X$ for the following PWNe (listed by increasing $\eta_X$): 
(1) Geminga,
(2) Vela,
(3) J1741--2054,
(4) B1951+32 (CTB 80)
(5) J1509--5850$^\dagger$,
(6) 3C 58, 
(7) B0355+54$^\dagger$,
(8) G11.2--0.3, 
(9) G54.1+0.3,
(10) B1509--58 (MSH 11--52), 
(11) G21.5--0.9, 
(12) Mouse (shown in red), 
(13) Crab$^\dagger$, and
(14) Kes 75.  The $\Gamma$ values for PWNe noted with $^\dagger$ were taken from Klingler et al.\ (2016a,b) and Mori et al.\ (2004), and all others were obtained using the spatially-resolved spectral map procedure described above (see Kargaltsev et al.\ 2017b for spectral maps of these PWNe).  The distances and parameters used to obtain the efficiencies were taken from Kargaltsev \& Pavlov (2010), and those for PSR J1741--2054 from Auchettl et al.\ (2015); we assume 30\% uncertainties for pulsar distances (except for B0355, for which a more precise parallax-based distance is known).}
\label{fig-eta}
\end{figure}

\begin{figure*}
\epsscale{1.2}
\plotone{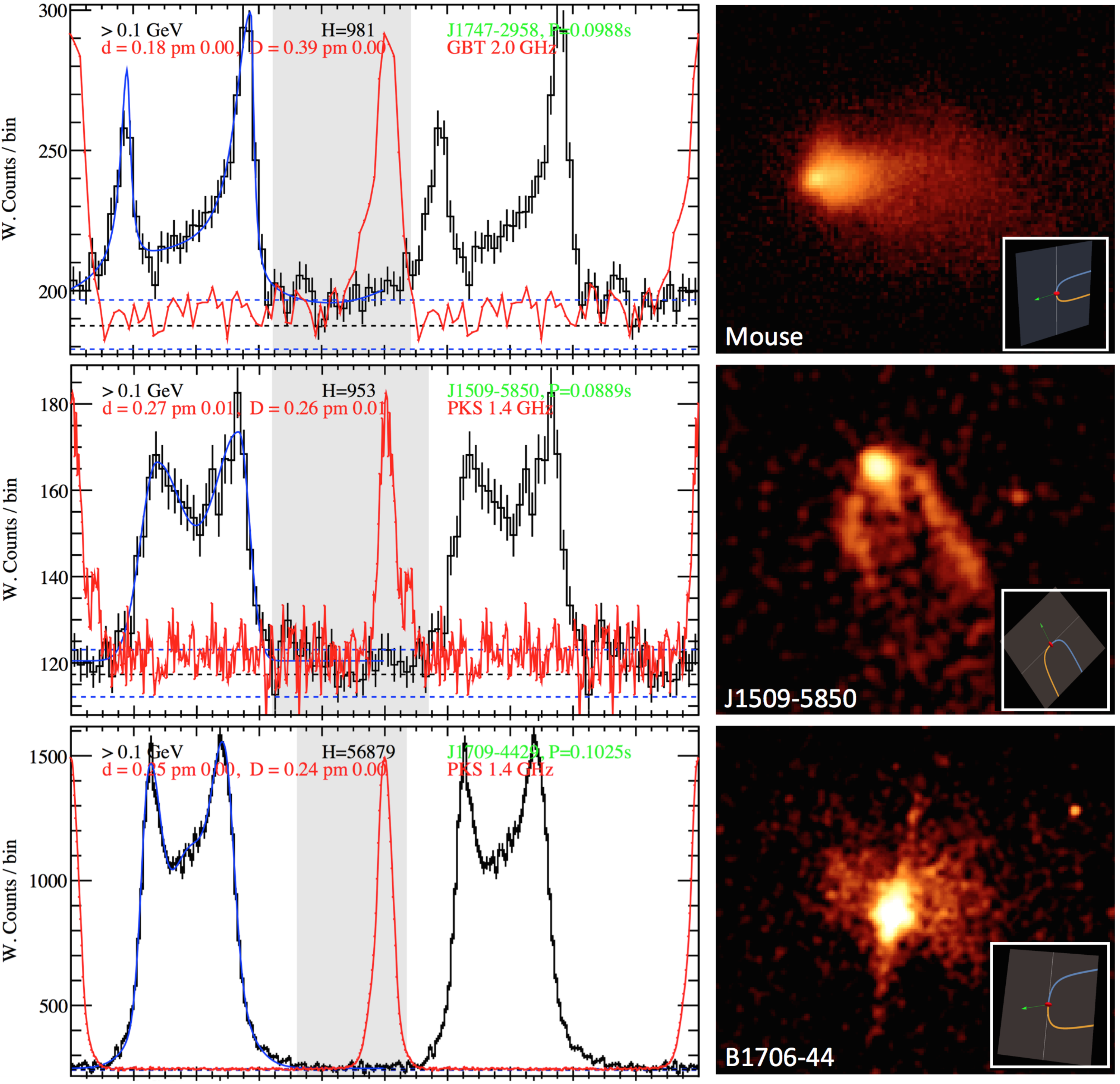}
\caption{Radio and $\gamma$-ray pulse profiles of the Mouse (top), J1509--5850 (middle), and B1706--44 (J1709--4429, bottom) pulsars, and {\sl Chandra} ACIS images of their PWNe.  The insets in the X-ray images are the schematic illustrations of the probable orientations of the spin axis, with the jets (initially launched along the spin axis) bent by the ram pressure due to pulsar motion (the motion direction is shown by green arrows).  The pulse profiles taken from the 2nd {\sl Fermi} LAT pulsar catalog (Abdo et al.\ 2013) show the weighted counts per bin as a function of rotational phase (2 periods are shown), and the H-test value $H$.  The letters $d$ and $D$ represent the lag of the first $\gamma$-ray peak relative to the radio fiducial phase, and the separation of the outermost $\gamma$-ray peaks, respectively
(see Section 5.1 and Appendix A of Abdo et al.\ 2013 for details).}
\label{fig-lightcurve}
\end{figure*}

The peak of the X-ray brightness in the ACIS image of the compact PWN is within $0\farcs35$ (0.7 ACIS pixels) from the radio timing position of the pulsar (corrected to the epoch of the ACIS observation), which is within the range of typical offsets in the {\sl Chandra} World Coordinate System\footnote{See http://cxc.harvard.edu/cal/ASPECT/celmon/}.  
In the HRC image, the shape of the brightness peak is coincident with a point source (see the right panel in Figure \ref{fig-images}). 
This indicates that the unresolved X-ray emission (as seen in the ACIS images) may be coming from the pulsar at the heart of Mouse PWN.  
The spectrum of the PWN core extracted from an $r=0\farcs74$ aperture (see Section 3.2.1) is nonthermal and can be described by an absorbed PL with $\Gamma=1.55\pm0.04$.  
This slope is only slightly smaller than $\Gamma=1.65\pm0.02$ obtained from the absorbed PL fit to the emission from the immediate vicinity of the pulsar (region 2 in Figure \ref{fig-regions}).  
Therefore, there is no spectral evidence for pulsar emission dominating the PWN emission in the core of the nebula. 
The flux extracted from the core region corresponds to an X-ray luminosity $L_{\rm X,psr}=(5.8\pm0.1)\times10^{33}$ erg s$^{-1}$ (in 0.5--8 keV, at $d=5$ kpc), yielding an X-ray efficiency $\eta_X \equiv L_X / 4\pi d^2=2.3\times10^{-3}$\footnote{Since the fraction of flux due to nebular emission in the vicinity of the pulsar is unknown, this value should be considered an upper limit.}. 
The core luminosity is a factor of 3.5 smaller than the luminosity of the entire PWN (excluding the pulsar) in X-rays, $L_{\rm X,pwn}=(2.03\pm0.02)\times10^{34}$ erg s$^{-1}$, with efficiency $\eta_X=8.1\times10^{-3}$. 
In Figure \ref{fig-eta} we compare $\eta_X$ and the spectral slope of the X-ray spectrum emitted by particles injected at the termination shock with those of other X-ray-bright PWNe. 
Assuming uncooled spectra (which is likely to hold in the vicinity of the pulsar), the observed spectral slopes, $1< \Gamma_i < 2$, correspond to slopes $1 < p_i < 3$ ($p_i\equiv2\Gamma_i-1$) of the particle SED, $dN/dE_e\propto E_e^{-p_i}$ ($p_i\approx2.3$ for the Mouse).

\subsection{Compact Nebula and Tail}

Hales et al.\ (2009) measured the transverse velocity of the Mouse PWN, $v_\perp = (306\pm43)d_5$ km s$^{-1}$.  
Together with the HRC image, this measurement allows us to estimate the Mach number for the pulsar. 
The HRC image (Figure 1, right panel) shows extended emission at the apparent apex of the PWN, resolved from the core, which is seen for up to $\simeq1''$ ahead of the pulsar.  
This corresponds to the projected distance $r_{\perp,s}=7.5\times10^{16}d_5$ cm.   
At this distance, the pulsar wind pressure, $P_{w} = \dot{E} f_\Omega (4\pi c r_s^2)^{-1} = 1.2\times 10^{-9} f_\Omega d_5^{-2}$ dyn cm$^{-2}$ ($f_\Omega$ takes into account PW anisotropy; $f_\Omega=1$ for isotropic wind), is balanced by the sum of the ambient pressure, $P_{\rm amb}= \rho kT (\mu m_{\rm H})^{-1}= 1.4\times 10^{-12} n T_4$ dyn cm$^{-2}$, and the ram pressure, $P_{\rm ram}= \rho v^2 = 1.7\times 10^{-10} n v_7^2=1.7\times 10^{-10} n v_{\perp,7}^2 \sin^{-2} i$ dyn cm$^{-2}$. 
Here $T=10^4 T_4\,{\rm K}$ is the ISM temperature, $v =10^7 v_7\,{\rm cm\, s}^{-1}$ is the pulsar velocity, $i$ is the angle between the line of sight and the pulsar motion direction, $\mu$ is the ISM mean molecular weight\footnote{We assume solar abundances, $\mu=1.37$, in all estimates.}, and $n = \rho/(\mu m_{\rm H})$ is the ISM number density in units of cm$^{-3}$. 
Assuming $P_{\rm ram}\gg P_{\rm amb}$ (or ${\mathcal M} \gg 1$), we estimated the ambient ISM number density as $n\simeq 0.8 f_\Omega \sin^{2} i$ cm$^{-3}$ (if the velocity vector is close to  the plane of sky, the shape of the bow shock in the vicinity of the apex can be approximated by a sphere resulting in $r_s\simeq r_{\perp,s}$).
The above estimate of the number density suggests a warm or hot ISM phase with the sound speed  $c_{\rm ISM}\simeq 10-100$ km s$^{-1}$  (see Section 21.1.2 of Cox 2000), unless $f_{\Omega}\gg 1$.  
We can also estimate the Mach number $\mathcal{M}= (P_{\rm ram}/\gamma_{\rm a} P_{\rm amb})^{1/2}= v/c_{\rm ISM}$, which typically ranges from a few to $\sim$30 for pulsars in the ISM (here $\gamma_{\rm a}=5/3$ is the ISM adiabatic index). 
Lower values of $\mathcal{M}$ (i.e., $\mathcal{M}\sim1$) are not likely as it would be difficult to explain the presence of the very long collimated radio tail.

In radio, the PWN behind the pulsar appears to expand laterally up to $\sim$1.5$'$ from the pulsar, while the X-ray PWN behaves quite differently: it steadily narrows beyond $\simeq 15''$ downstream from the pulsar (see Figures 1 and 3). 
The narrowing of the X-ray PWN could be explained by more efficient cooling of X-ray emitting electrons located further from the tail's axis caused by either a substantially slower flow velocity compared to the inner channel of the tail, or by a stronger magnetic field. 
The latter is not likely because the higher magnetic field would enhance the synchrotron emission brightness in radio and in X-rays further from the tail's axis (which is not seen) unless the electron density is lower there. 
On the other hand, the former interpretation contradicts to the results of numerical simulations (see Figure~1 of Bucciantini et al.\ 2005) which suggest that the flow is slower in the inner channel behind the pulsar. 
This, however, may be the result of limitations in the simulation setup.
The lower energy radio-emitting electrons have much longer cooling timescales and therefore their appearance in the images is not affected across distances on the order of the size of the X-ray nebula.

\begin{figure*}
\epsscale{1.15}
\plotone{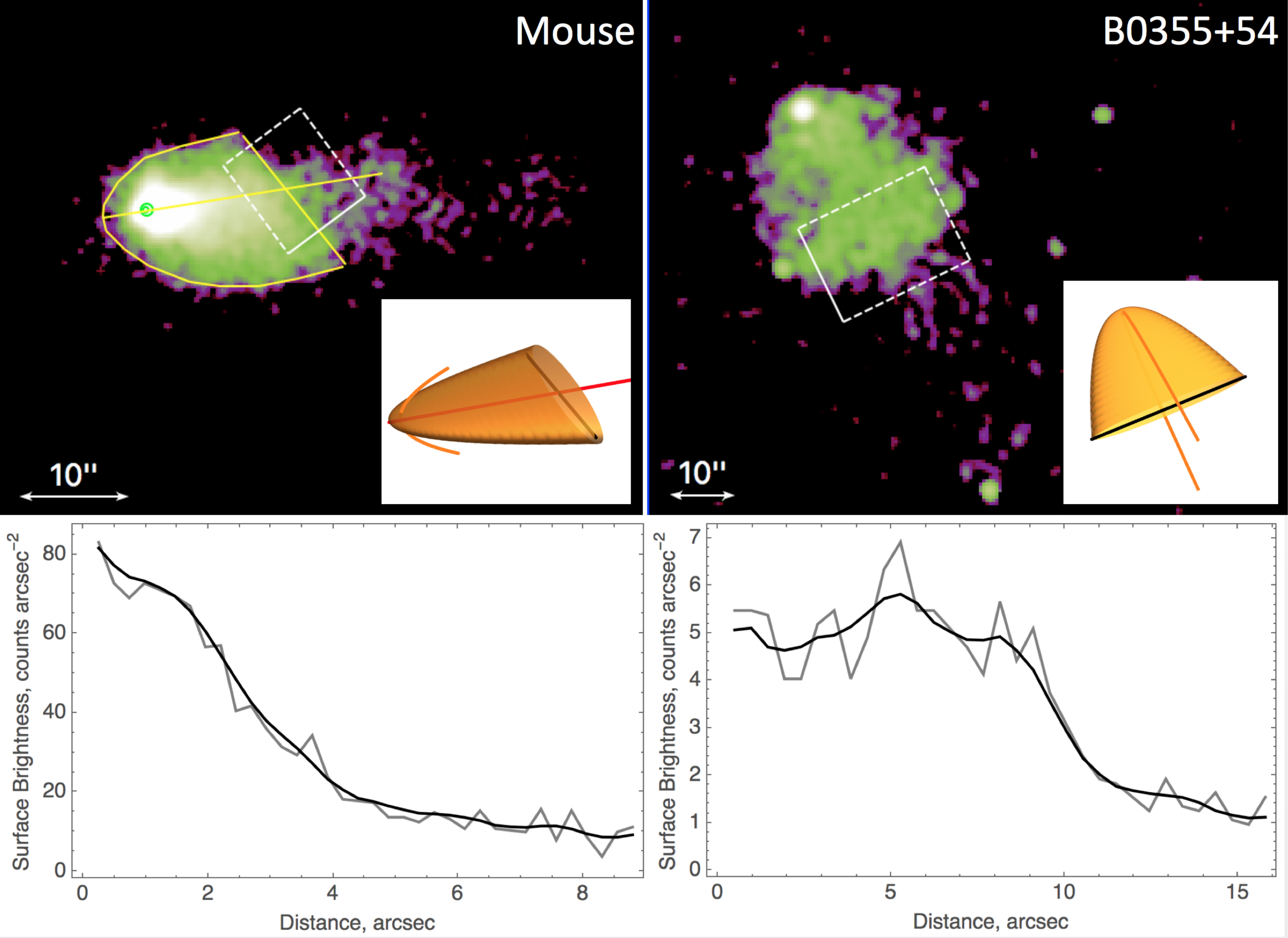}
\caption{{\sl Top Left:} Merged ACIS image of the Mouse CN, binned by a factor of 0.5, and smoothed with a 3-pixel ($r=0\farcs74$) Gaussian kernel.  The solid yellow contour outlines the outer edge of the putative deformed equatorial outflow (see Section 4.3 for discussion), and the yellow line passing through the pulsar and trailing edge of the CN shows the symmetry axis of the CN (assuming the CN resembles the 3D structure shown in the inset).  
Shown in the inset is a 3D illustration of pulsar wind flattened near the CN equatorial plane, viewed from the same angle as the Mouse ($\sim20^\circ$ offset from the equatorial plane).  The curved orange lines illustrate polar outflows (jets) bent back by the ram pressure.
{\sl Bottom Left:} Plot of the surface brightness of the downstream end of the CN, obtained from the region enclosed by the dashed white box shown in the top left image.  The gray curve was obtained without smoothing applied, and the black curve was obtained with the smoothed image.
{\sl Top Right:}  ACIS image of the B0355+54 CN (smoothed with an $r=1\farcs48$ Gaussian kernel), which features a similar CN composed of a deformed equatorial outflow and a similar trailing edge, viewed from near the pulsar spin (i.e., jet) axis (see Klingler et al.\ 2016b), with a similar inset, shown for comparison.
{\sl Bottom Right:}  The surface brightness profile for the B0355 CN plotted in the same manner.}
\label{fig-3d}
\end{figure*}

In the X-ray images, the Mouse CN displays a ``filled'' morphology similar to those of the X-ray CNe around the supersonic B0355+54 (Klingler et al.\ 2016b) and J1741--2054 (Auchettl et al.\ 2015) pulsars. 
Such a morphology is in contrast to the ``hollow'' morphologies of the X-ray CNe of Geminga (Posselt et al.\ 2017) and J1509--5850 (Klingler et al.\ 2016a), which are also moving supersonically. 
A somewhat abrupt decrease in brightness occurring $\simeq10''$-$15''$ behind the pulsar is shown by the dashed line in Figure 1 (left panel).
The drop in surface brightness is also seen in the VLA 1.5 GHz image, at the same position and angle as seen in the X-rays. 
A similarly abrupt drop in X-ray surface brightness occurs in the B0355+54 CN at a comparable distance behind the pulsar (Klingler et al.\ 2016b; see Figure \ref{fig-3d}, top right panel).  
However, in the B0355 CN, which appears to be symmetric with respect to the pulsar's motion trajectory, the drop occurs along a line perpendicular to the motion direction.  
For the Mouse CN, this line is inclined substantially with respect to the Mouse pulsar velocity direction (see Figure \ref{fig-3d}, top left panel). 
If this line corresponds to the edge of the same type of flattened structure as seen in the B0355+54 PWN (e.g., an abrupt change in the properties of an equatorial outflow deformed by the ram pressure), then its apparent lack of perpendicularity with respect to the proper motion direction is due to projection effects and the 3D orientation of the system.
The projected inclination of the downstream brightness drop implies that the equatorial plane is inclined $20^\circ$ from our line of sight, corresponding to a viewing angle $\zeta\sim70^\circ$ from the pulsar spin axis.
This also implies that a small component of the Mouse's velocity vector is pointing out of the plane of the sky toward the observer.
The B0355+54 CN appearance can be explained by a similar equatorial outflow if the pulsar's velocity vector is perpendicular to the line-of-sight and the equatorial plane is close to the plane of the sky (Figure \ref{fig-3d}, top right inset).  
The above interpretation of the Mouse PWN's orientation (i.e., the velocity vector is nearly perpendicular to the spin axis, and $\zeta\sim70^\circ$) is compatible with the viewing geometry suggested by the $\gamma$-ray light curve (see above), which also indicates that our line of sight is oriented near the equatorial plane (see Section 4.1).
Since radio pulsations are seen, this also supports the interpretation that the pulsar's magnetic axis is significantly offset from its spin axis ($\alpha\sim70^\circ$).

Despite the above-discussed similarities between the CNe of the Mouse and B0355, the extended tails of these two pulsars show noticeable differences.   
The B0355 PWN spectra show no signs of spectral softening between the CN and its long X-ray tail, nor along the tail up to distances $\sim2$ pc, while the Mouse PWN shows clear evidence of fast softening on scales $<0.1$ pc (see Figure \ref{fig-cooling}). 
The differences could be explained if the Mouse tail has either a much higher magnetic field strength or a much slower bulk flow velocity. 
The latter seems to be difficult to reconcile with the extremely long (36 pc) collimated radio tail (see Figure 1 in Yusef-Zadeh \& Gaensler 2005).
Therefore, the former interpretation is more likely unless there is substantial re-acceleration of the flow at large distances from the pulsar in the Mouse's tail or if some ``re-heating'' mechanism operates in the tail of B0355 which does not operate in the Mouse tail (e.g., the flow is much more turbulent in the B0355 tail).
In addition, the B0355 tail is far less collimated compared to that of the Mouse, which may explain why it is not seen in radio (Ng et al., in prep.). 
The reduced collimation is likely the consequence of a lower pulsar velocity ($v_{\rm B0355}=61^{+12}_{-9}$ km s$^{-1}$; Chatterjee et al.\ 2004).

Finally, the 4.89 GHz VLA image of the Mouse tail (Figure 3, top right panel) shows hints of quasi-regular brightness enhancements in the faint narrow part of the radio tail at $\gtrsim 3'$ from the pulsar. 
Unless these enhancements are due to instrumental artifacts\footnote{In radio interferometer images, a ripple-like artifact pattern often surrounds bright sources and is usually seen in the faint background.  A faint striped artifact pattern is seen in the background of this image, but its scale and orientation are not consistent with the brightness enhancements seen in the tail.}, they may reflect the structure of magnetic field in the tail (e.g., magnetic field enhancements possibly associated with multiple Mach disks; see Morlino et al.\ 2015), but further observations are needed to confirm their presence.

\subsection{Multi-Wavelength Spectrum}
\label{sec:MWspec}

\begin{figure*}
\epsscale{1.15}
\plotone{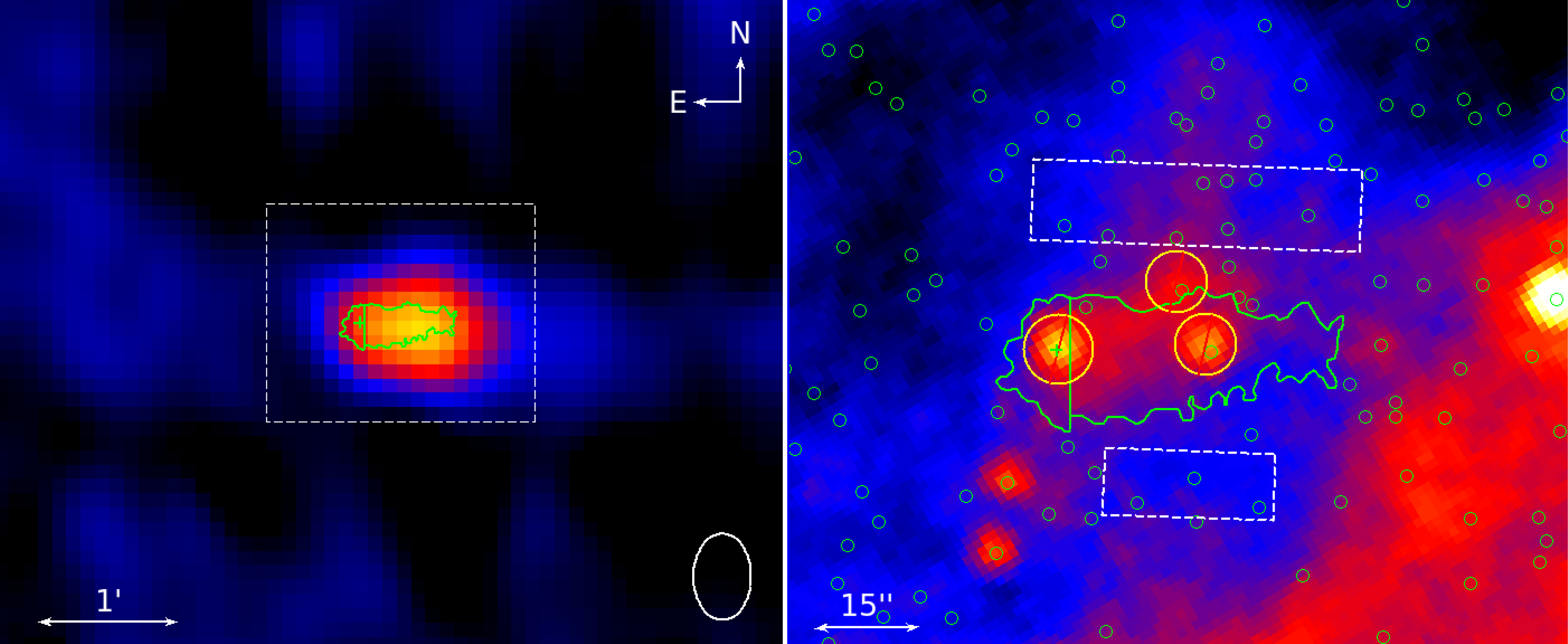}
\caption{{\sl Left:} GMRT 150 MHz image of the Mouse PWN.  The white ellipse in the bottom right corner shows the synthesized beam size, and the dotted white box shows the field of view shown in the right panel. {\sl Right:} {\sl Spitzer} MIPS 24 micron image of the Mouse field.  The green circles mark nearby 2MASS point sources.  In both of the panels the green cross marks the pulsar position, and the X-ray tail contour from Figure 3 is shown.  The yellow circles encompassing the 2MASS sources detected by {\sl Spitzer} were excluded from the estimation of the upper limit of the tail's IR flux, and the dashed white boxes were used for background subtraction.}
\label{fig-gmrt-spitzer}
\end{figure*}

The Mouse has been previously observed with the VLA between 1.5 and 15 GHz (Camilo et al.\ 2002; Yusef-Zadeh \& Gaensler 2005; Hales et al.\ 2009) and has been identified in a 330 MHz VLA observation (Hyman et al.\ 2005).  
We have also investigated the 150 MHz GMRT pipeline-produced image from the TIFR GMRT Sky Survey (TGSS; Intema et al.\ 2017) and the {\sl Spitzer} image from the 24 ${\rm \mu}$m MIPS Galactic Plane Survey (Gutermuth \& Heyer 2015), which are shown in Figure \ref{fig-gmrt-spitzer}.  
In both cases we found extended objects positionally coincident with the Mouse CN. 
In the low-resolution GMRT image (Figure \ref{fig-gmrt-spitzer}, left panel) the source has an elongated shape consistent with the extent of the X-ray tail.   
In the much higher resolution 24 ${\rm \mu}$m image, most of the emission comes from the vicinity of pulsar position (associated with the source MG359.3052--00.8412; Gutermuth \& Heyer 2015). 
The lack of 2MASS counterpart hints that MG359.3052--00.8412, with a flux density of $17.7 \pm 2.1$ mJy at $\lambda=24\,\mu{\rm m}$, may be associated with the Mouse PWN head. 
However, even in this case, it might not be associated with the pulsar wind synchrotron emission\footnote{For instance, the Crab PWN spectrum has a bump in IR due to emission from interstellar dust (Atoyan \& Aharonian 1996).}.  
Fainter extended emission is seen extending west of MG359.3052--00.8412, but it does not conform well to the shape of the X-ray or radio tail, so it could be due to (or heavily contaminated by) unrelated foreground/background emission. 
The 24 ${\rm \mu}$m flux of this emission is difficult to accurately measure due to the complex nonuniform background emission in the vicinity (and the flux is highly dependent upon background treatment), but we obtain a conservative upper limit of $1.3\times10^{-11}$ erg s$^{-1}$ by subtracting the background from the regions shown in Figure 10. 
High-resolution IR observations are needed (with the {\sl HST} or {\sl JWST}) to determine the nature of this emission and its connection to the Mouse PWN.

\begin{figure*}
\epsscale{1.1}
\plotone{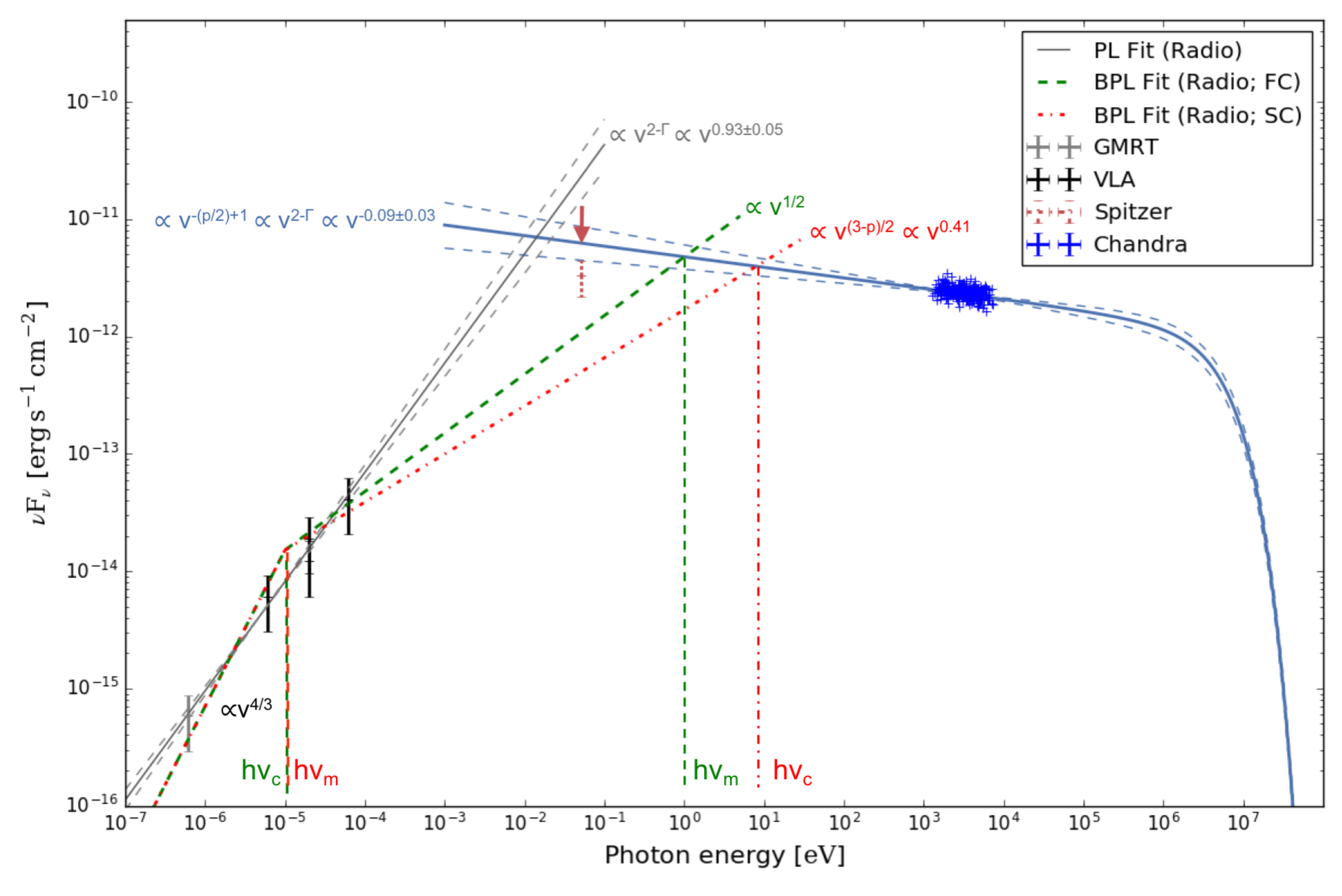}
\caption{Multiwavelength spectrum of the X-ray tail region (the right section of the contour shown in Figure \ref{fig-radio}, left panel; this region excludes emission from the pulsar).  
The following data points are plotted:  GMRT: 150 MHz;  VLA: 1.47 GHz, 4.77 GHz (2 different array configurations), 14.9 GHz;  {\sl Spitzer}: 24 $\mu$m -- the tail flux upper limit shown is by the arrow, the bright CN head (MG359.3052--00.8412) is shown by the dashed point; and the unabsorbed 1.2--8 keV {\sl Chandra} fluxes (there were insufficient counts below 1.2 keV due to absorption).  
The high-energy synchrotron spectrum (blue lines) extrapolated from the {\sl Chandra} data (blue points) was plotted with {\tt naima} assuming a PL SED for the electrons with $p_X=2\Gamma_X-2=2.18$ (obtained from fitting the X-ray data; $\Gamma_{\rm X}=2.09\pm0.03$). 
The magnetic field $B=250$ $\mu$G (the choice of this value is explained in Section 4.5), and the exponential cutoff energy of the electron SED was set at $3.35$ PeV, which is the maximum accelerating potential of the pulsar's polar cap, $\Phi=(3\dot{E}/2c)^{1/2}$ (however, the spectrum is likely to exhibit an exponential cutoff at frequencies much lower than those corresponding to the maximum accelerated particle energy, and hence this should be interpreted as an upper limit for the cutoff energy). 
The grey lines show a PL fitted to the radio data points ($\Gamma_{\rm R}=1.07\pm0.05$).  
Various spectral slopes corresponding to low-energy segments of PLs with double breaks correspond to the cooling regimes discussed in the text: $\nu_c < \nu_m$ -- green, $\nu_m < \nu_c$ -- red; the vertical lines are shown for illustrative purposes.
The dashed lines (gray and blue) represent $1\sigma$ uncertainties of the fitted slopes.}
\label{fig-mw-spectra}
\end{figure*}

The radio and X-ray measurements (obtained using the X-ray tail region shown in the left panel of Figure 3) suggest that the multiwavelength spectrum from this region (see Figure \ref{fig-mw-spectra}) should experience at least one break between $\sim10^{-5}$ eV ($\sim$ 2 GHz) and 500 eV if the spectrum can be approximated by a combination of PLs (see Figure \ref{fig-mw-spectra}).

If a single break (around $10^{-2}$ eV) is assumed (illustrated by the gray line in Figure \ref{fig-mw-spectra}), then the low-frequency radio spectrum can be fitted\footnote{When fitting a PL to the radio fluxes, we weighted the values by the inverse of the corresponding fractional uncertainties squared. For the GMRT flux measurement, the extraction region size is comparable to the beam size, so we assume a 50\% fractional uncertainty.  For the VLA data, the precise values of uncertainties depend on the specifics of the data reduction pipeline parameters used and hence difficult to estimate.  Since at 4.77 GHz two different array configurations lead to fluxes that differ by roughly 50\%, we also conservatively ascribe a 50\% uncertainty to all VLA flux measurements.} 
by a PL with $\Gamma_{\rm R}=1.07\pm0.05$ with a low-frequency (150 MHz to 0.01 eV) luminosity of $(7.9\pm3.4)\times10^{31}$ erg s$^{-1}$ and a high-frequency (0.01 eV to 10 MeV) luminosity of $2\times10^{35}$ erg s$^{-1}$ for the X-ray tail region. 
The estimated luminosity is only weakly dependent on the choice of upper limit, so we pick an arbitrary but plausible value of 10 MeV (the energy to which the Vela PWN's spectrum extends to); using an upper limit of 100 keV results in roughly a 10\% decrease in luminosity.
The magnitude of the spectral break between the radio and X-ray spectra\footnote{We note that although the spatially-averaged {\sl CXO} spectrum extracted from the X-ray tail region fits a PL with $\Gamma_{\rm X}=2.09\pm0.03$, the spatially-resolved spectroscopy shows that the slope depends on distance from the pulsar (see Section 3.2.2).} corresponds to $\Delta\Gamma = \Gamma_X - \Gamma_R \approx1.0\pm0.1$. 
Although large ($>$0.5) values of $\Delta \Gamma$ between the radio and X-ray spectra are common for PWNe (e.g., Chevalier 2005; Reynolds et al.\ 2017), they are challenging to obtain from standard synchrotron cooling considerations
(which predicts $\Delta\Gamma=0.5$) unless additional mechanisms, such as entrainment, turbulent magnetic field amplification, diffusion, and/or reconnection are invoked. 
The modeling of these processes is beyond the scope of our paper; below we briefly discuss synchrotron cooling scenarios and the physical constraints they impose.

Although the existing data are consistent with a single break in the PL spectrum, a double break (or more complex behavior) is not excluded. 
In fact, one would expect to see a double break in a spatially-integrated synchrotron spectrum of a cooling outflow if the observed spectrum includes frequencies lower than the characteristic synchrotron frequency corresponding to the lower boundary of the tail's electron SED (see Appendix A).
If the injected electron SED is a single PL (with slope $p_i$) between lower and upper boundary energies $\gamma_m m c^2$ and $\gamma_M m c^2$, respectively, then cooling will affect the spectrum if $\nu_c \lesssim \nu_M=\nu_{\rm cyc}\gamma_M^2$, and the breaks occur near the energies 
\begin{equation}
h\nu_m =h\nu_{\rm cyc}\gamma_m^2 = 1.2\times 10^{-6} \left(\frac{B}{100\,\mu{\rm G}}\right) \left(\frac{\gamma_m}{1000}\right)^2\,\, {\rm eV} 
\label{hnum}
\end{equation} 
and 
\begin{equation}
h\nu_c = h\nu_{\rm cyc}\gamma_c^2
= 7 \left(\frac{100\,\mu{\rm G}}{B}\right)^3 \left(\frac{1\,{\rm kyr}}{t_{\rm tail}}\right)^2\,\,{\rm eV}\,.
\label{hnuc}
\end{equation}
In these equations $\nu_{\rm cyc}=eB/(2\pi mc)=2.8 B$ MHz is the (nonrelativistic) cyclotron frequency, $\gamma_c=t_{\rm cyc}/t_{\rm tail}$ is the Lorentz factor at which the synchrotron cooling time equals the maximum residence time $t_{\rm tail}$ of the pulsar wind electrons in the tail region, and $t_{\rm cyc} =9m^3c^5/(4e^4B^2)= 7.74\times 10^8 B^{-2}$ s is the cyclotron cooling time (see Appendix A).
Since $\gamma_m$ is not known, the ordering of $h\nu_c$ and $h\nu_m$ is also not known (however, they can be estimated and their ordering can be determined from other constraints, which we discuss in Section 4.5). 
Following Sari et al.\ (1998), we will define the cases $\nu_c > \nu_m$ and $\nu_c < \nu_m$ as the {\it slow cooling} and {\it fast cooling} regimes, respectively. 
The spectral flux densities $F_\nu$ in the two regimes are given by Equations (A1) and (A2) (see also Sari et al.\ 1998 and Section 5 of Piran 2004).
Note that in both regimes $F_\nu\propto \nu^{1/3}$ in the low-frequency part of the spectrum, at $\nu\ll \nu_m\nu_c/(\nu_m+\nu_c) \approx {\rm min}(\nu_m,\nu_c)$, while $F_\nu\propto \nu^{-p_i/2}$ in the high-frequency (cooled) part of the spectrum, at $\nu\gg {\rm max}(\nu_m,\nu_c)$.

Possible multiwavelength spectra of the Mouse tail with two breaks are shown in Figure \ref{fig-mw-spectra} (the red dashed-dotted line implies $\nu_m < \nu_c$ while the green dashed-dotted line implies $\nu_c < \nu_m$).
Since the observed X-ray spectrum is softening rapidly along the tail, it is reasonable to assume that the X-ray emission is in the cooled part of the spectrum.
In this case, one can relate the slope $p_i$ of the injected particle SED to the X-ray photon index as $p_i=2\Gamma_X-2$, regardless of whether $\nu_m > \nu_c$ or vice versa. 
This leads to a more conventional $p_i \approx2.2$ expected from acceleration in relativistic shocks (see, e.g., Bykov et al.\ 2017 and references therein) compared to $p_i \approx3.2$ which would follow from $p_i = 2\Gamma_X -1$ for the uncooled part of spectrum. 
Note that the same injected SED slope, $p_i = 2\Gamma_i -1 = 2.2$ (for $\Gamma_i = 1.6$; see Figure 5 and Table 3) was obtained from the PL fit of the tail in the immediate vicinity of the pulsar.

In principle, an additional spectral break could be expected at very low frequencies, where the radiating volume becomes optically thick with respect to absorption of synchrotron radiation, which leads to suppression of the synchrotron spectrum. 
Assuming a known flux density $F_{\nu, \rm obs}$ at some observed frequency $\nu_{\rm obs}$ below $\nu_m$ and below $\nu_c$, the spectrum becomes self-absorbed at a frequency $\nu_{SA}$ which can be estimated from (see, e.g., Beniamini $\&$ Kumar 2016)
\begin{equation}
\label{eq:nuSA}
\frac{2 \nu_{\rm SA}^2}{c^2}\gamma(\nu_{\rm SA})m c^2 \frac{A}{4\pi d^2}=\bigg( \frac{\nu_{\rm SA}}{\nu_{\rm obs}}\bigg)^{1/3}F_{\nu, \rm obs}
\end{equation}
where $\gamma(\nu_{\rm SA})=(2\pi m c \nu_{\rm SA}/eB)^{1/2}$ is the Lorentz factor of electrons radiating at $\nu_{\rm SA}$, and $A$ is the emitting surface area.
We use the surface area of the assumed emitting cylindrical volume with radius $r=7''$ and length $l=38''$ described above.
The specific values of the self-absorption frequency are not very sensitive to the exact choice of geometry (which would slightly change the estimate of surface area).
For $\nu_{\rm obs}$ and $F_{\rm \nu,obs}$, we take the lowest available observed frequency (and corresponding flux) of 150 MHz because we see that, at least down to those frequencies, the spectrum has not been sufficiently steepened, so self-absorption must be below the observed range.
We also choose a low enough $\nu_{\rm obs}$ so that we can be sure that $F_\nu$ scales as $\nu^{1/3}$ below that frequency (which holds for either the globally slow or globally fast cooling regimes at low enough frequencies).
Solving Equation (\ref{eq:nuSA}) we find $\nu_{\rm SA}\simeq 60$ kHz.
The self-absorption frequency is therefore expected to lie far below the observed range, and should not affect the observed spectrum.
For illustrative purposes, even if we take $\nu_{\rm obs}$ at 1.5 GHz (the lowest frequency observed with the VLA), the estimate of $\nu_{\rm SA}$ changes by less than 20\%.
Thus, since we find $\nu_{\rm SA}$ to be orders of magnitude below the observed range of frequencies, the choice of $\nu_{\rm obs}$ (and $F_{\rm \nu,obs}$) is not a critical one.

If there are two breaks in the Mouse tail's spectrum, the overall change in slope (from radio to X-rays) is $\Delta\Gamma = 0.77\pm0.05$ because the lower energy spectrum is $F_\nu \propto \nu^{1/3}$, regardless of whether $\nu_m>\nu_c$ or vice versa.
This change is consistent with the existing measurements within their uncertainties and is somewhat smaller than $\Delta\Gamma\approx1.0\pm0.1$ for a PL with a single break (see Figure \ref{fig-mw-spectra}). 
In order to firmly discriminate between the double break and single break PL scenarios, the Mouse PWN must be resolved at mm (with ALMA) and NIR (with the {\sl HST} or {\sl JWST}) wavelengths.

\subsection{Magnetic Field, Energetics, and Flow Velocity}

The spectra extracted from the X-ray tail can be used to estimate the magnetic field. 
If the X-ray emission is in the cooled part of spectrum ($\nu > \nu_c$), the commonly used equations for the uncooled single PL spectrum (e.g., Equation (7.14) in Pacholczyk 1970 or a more general Equation (2) in Klingler et al.\ 2016a) are inapplicable. 
Therefore, in Appendix B we derive new equations with allowance for synchrotron cooling for both the slow cooling and fast cooling regimes (Equations (B8) and (B11), respectively) and use these equations for magnetic field estimates in the Mouse's tail.

The magnetic fields estimated from Equations (B8) and (B11) are proportional to $[L(\nu_1,\nu_2)/V]^{2/7}$, where $L(\nu_1,\nu_2)$ is the luminosity in the frequency range $\nu_1<\nu<\nu_2$ and $V$ is the radiating volume. 
For $\nu_1$ and $\nu_2$ we use the frequencies corresponding to 0.5 and 8 keV, the range over which $L(\nu_1,\nu_2)\approx 2\times 10^{34}d_5^2$ erg s$^{-1}$ and $\Gamma\approx 2.1$ have been measured (see Table 3). 
As $L(\nu_1,\nu_2)\propto d^2$ and $V\propto d^3$, the magnetic field estimate is not very sensitive to the assumed distance, $B\propto d^{-2/7}$. 
For a cylindrical synchrotron-emitting region of $7''$ radius and $38''$ length (the approximate dimensions of the tail region; see Figure 3), located at $d=5$ kpc, the volume is $V\approx2.4\times10^{54}$ cm$^3$.
The right-hand-sides of Equations (B8) and (B11) also depend on $\nu_c$, $\nu_m$, and $\nu_M$. 
The estimated $B$ value is, however, insensitive to $\nu_M$ as long as $(\nu_c/\nu_M)^{\Gamma -1.5}\ll 1$ (which holds when cooling significantly affects the spectrum). 
The other two frequencies can be crudely estimated from the expected values of the break frequencies in the observed multiwavelength spectrum (see Figure 11).
Varying these frequencies in plausible ranges and the spectral index within its uncertainties, we obtained equipartition magnetic fields $B_{\rm eq} \equiv B\sigma^{-2/7} \sim(170$--250) $\mu$G for the slow cooling regime, and $B_{\rm eq}\sim(100$--110) $\mu$G for the fast cooling regime, where $\sigma = w_B/w_e$ is the magnetization parameter ($w_B$ and $w_e$ are the energy densities of the magnetic field and relativistic electrons). 
These magnetic fields are higher than typical $B_{\rm eq}\sim (10$--100) $\mu$G estimates for most PWNe. 
They become even higher if the emitting volume is smaller, due to a possibly flattened tail structure (see Section 4.3).

Using the magnetic field estimates, we can estimate the tail's energy transfer (injection) rate (see Klingler et al.\ 2016a), 
\begin{eqnarray}
\dot{E}_{\rm tail} & \sim & (w+p) S v_{\rm flow} 
\label{energy_injection}\\
\nonumber
& \sim & 2.9\times 10^{35} (B_{\rm eq}/200\,\mu{\rm G})^2 (S/7.5\times 10^{35}\, {\rm cm}^2)\\
\nonumber
&\times & (v_{\rm flow}/1000\,{\rm km}\,{\rm s}^{-1}) [\sigma^{4/7}+(2/3)\sigma^{-3/7}]^2\,\,{\rm erg}\,{\rm s}^{-1} ,
\end{eqnarray}
and energy density
\begin{eqnarray}
w & = & w_B + w_e = [B_{\rm eq}^2/(8\pi)](\sigma^{4/7}+\sigma^{-3/7})^2 \\
\nonumber
 & \sim & 1.6\times 10^{-9} (B_{\rm eq}/200\,\mu{\rm G})^2 (\sigma^{4/7}+\sigma^{-3/7})^2\,\,{\rm erg}\,{\rm cm}^{-3}.
\label{energy_density}
\end{eqnarray}
Here, $v_{\rm flow}$ is the average velocity of the flow passing through cross-sectional area $S$, $w+p$ is the enthalpy density, and $p=w_B + w_e/3$ is the pressure of the relativistic particles and the magnetic field. 
When $\sigma=3/4$, the term $(\sigma^{4/7}+\sigma^{-3/7})^2\approx3.92$ and is minimized. 
Equation (\ref{energy_injection}) and the requirement that $\dot{E}_{\rm tail} < \dot{E}$ can be used to constrain $v_{\rm flow}$. 
If the outflow is conical with a width of 13$''$ in the bright CN ($S=7.5\times10^{35}$ cm$^2$), then $v_{\rm flow}$ should be lower than $\approx 2200 (B_{\rm eq}/200\,\mu{\rm G})^{-2}$ km s$^{-1}$. 
However, if most of the X-ray tail is composed of a deformed equatorial outflow, then, at least initially, the tail may be best approximated as a flattened structure rather than a conical structure. 
This would reduce the cross-sectional area in Equation (4) and allow higher flow speeds. 
For example, for an outflow 15$''$ wide (the width of the CN at the brightness drop, see Figure \ref{fig-3d}) and 3$''$ thick (a plausible estimate), the upper limit on the bulk flow velocity could be as high $\sim10,000$ km s$^{-1}$.
The lower limit on the average flow velocity, $v_{\rm flow} > l_{\rm rad}/\tau_{\rm sd} \sim 600$ km s$^{-1}$, where $l_{\rm rad}\sim 17$ pc is the length of the radio tail, $\tau_{\rm sd}=26$ kyr is the spin-down age of the pulsar.

Since we can estimate the cooling break energy $h\nu_c$ from the observed multiwavelength spectrum, we can place additional constraints on $v_{\rm flow}$ and distinguish between the different interpretations of the multiwavelength spectrum. 
Substituting $t_{\rm tail} =l_{\rm tail}/v_{\rm flow}$ in Equation (\ref{hnuc}), we obtain
\begin{eqnarray}
\label{vflow}
v_{\rm flow} & =& \frac{l_{\rm tail}}{t_{\rm cyc}}\left(\frac{\nu_c}{\nu_{\rm cyc}}\right)^{1/2} \\
\nonumber
& =& 
1050 \left(\frac{B}{200\,\mu{G}}\right)^{3/2}
\left(\frac{h\nu_c}{1\,{\rm eV}}\right)^{1/2}
\left(\frac{l_{\rm tail}}{1\,{\rm pc}}\right)\,\,\frac{{\rm km}}{{\rm s}} 
\end{eqnarray}
($l_{\rm tail} = 1$ pc corresponds to the angular size of $42''$ at $d=5$ kpc, if the tail is oriented in the plane of the sky).

From the fast cooling ($\gamma_c \ll \gamma_m$) fit in the multiwavelength spectrum (Figure 11), we see that $h\nu_c\sim10^{-5}$ eV, which gives $v_{\rm flow}\sim 3 (B/200\,\mu{\rm G})^{3/2}$ km s$^{-1}$. 
Since such a low flow speed is clearly incompatible with the length of the radio tail (see above), we can dismiss the fast cooling scenario as unphysical.

For the slow cooling ($\gamma_m \ll \gamma_c$) fit, we have $h\nu_c\sim10$ eV, which gives $v_{\rm flow}\sim 4400$ km s$^{-1}$ (for $B=250$ $\mu$G) -- a reasonable bulk flow speed for a collimated pulsar wind.
Such a speed is compatible with the tail's energy injection rate and a flattened outflow geometry. 
It corresponds to the residence time $t_{\rm tail}=l_{\rm tail}/v_{\rm flow}\sim 200$ yr (for the X-ray tail), much smaller than the pulsar's spin-down age of 26 kyr. 
The slow cooling (high velocity) interpretation implies $h\nu_m\sim10^{-5}$ eV and $\gamma_m\sim 2000$, much lower than the $\gamma_M = \gamma_m (\nu_M/\nu_m)^{1/2} \sim 10^8$ for $h\nu_M\sim 100$ keV.
This indicates that a broad SED, spanning $\gtrsim 5$ orders of magnitude, is produced by the particle acceleration process.

\section{Summary}

We have analyzed new deep {\sl CXO} ACIS observations of the Mouse PWN, together with the archival {\sl CXO} ACIS and HRC observations, and explored archival low-frequency data on this remarkable object.
The deep X-ray observations show that the Mouse PWN consists of a filled CN, followed by a $45''$ long tail which narrows with distance from the pulsar.
In radio, the nebula expands laterally up to much larger distances ($\sim 1\farcm5$ behind the moving pulsar) and then forms a very long, $\sim 12'$ collimated tail.
In the HRC image we observe the projected bow shock apex $\approx$1$''$ ahead of the pulsar position.

The deep {\sl CXO} observations have produced a clearer picture of the PWN, which allows us to gain insight into the magnetic inclination and viewing angles of the pulsar when we also consider the geometry information encoded in the pulsar light curves. 
The observed PWN morphology can be interpreted as an equatorial outflow deformed by the ram pressure with the equatorial plane being inclined $\sim 20^\circ$ from the line-of-sight, which implies a viewing angle with respect to the pulsar spin axis $\zeta\sim70^\circ$ (consistent with the $\gamma$-ray pulse profile), and that a small component of the velocity vector is oriented toward the observer. 
From the presence of the radio pulsations, we can infer that the magnetic axis of the pulsar is inclined by $\alpha\sim$70$^\circ$ from its spin axis.

We find that the spectrum extracted from the pulsar region is well fit by an absorbed PL model with photon index $\Gamma=1.55\pm0.04$ and $N_{\rm H}=2.7\times10^{22}$ cm$^{-2}$, although this spectrum is most likely contaminated by the bright surrounding nebular emission or even dominated by this emission.  
The spectrum of the PWN softens with increasing distance from the pulsar, with photon index ranging from $\Gamma=1.65\pm0.02$ in the inner regions to $\Gamma\approx3.0\pm0.2$ at the end of the X-ray tail.
The spectral softening along the tail can be attributed to rapid synchrotron cooling in a strong magnetic field.  
We have also produced an adaptively-binned spectral map of the PWN to visualize the spectrum and the cooling trend.

At the distance of 5 kpc, the entire PWN exhibits a 0.5--8 keV luminosity $L_X \approx 2.0 \times10^{34}$ erg s$^{-1}$, yielding a relatively high X-ray PWN efficiency, $\eta_X\approx8.1\times10^{-3}$.
We found that the spatially-averaged X-ray spectrum of the entire X-ray tail can be well fit by a single absorbed PL with $\Gamma_{\rm X}=2.09\pm0.02$ (this highlights the problem with inferring the SED of particles injected at the termination shock from spatially-averaged spectra of large regions). 
We find the slope $p_i \approx 2.2$ of the electron SED injected at the termination shock, from both the spatially-resolved X-ray spectroscopy and the spatially-integrated tail spectrum.

We have identified the Mouse PWN in the 150 MHz image from the TIFR GMRT Sky Survey. 
We performed an assessment of the multiwavelength spectrum of the X-ray tail using the GMRT and VLA data and found that the shape of the spectrum between 150 MHz and 8 keV can be approximated by a broken PL with at least one break between $2.4\times10^{9}$ Hz ($10^{-5}$ eV) and $1.2\times10^{17}$ Hz (0.5 keV). 
Using the multiwavelength data, we estimated the equipartition magnetic field in the X-ray tail $B_{\rm eq}\sim200$ $\mu$G, which is consistent with the rapid synchrotron cooling observed. 
From the tail flow velocity constraints, we conclude that the multiwavelength spectrum likely has two breaks -- the cooling break at $h\nu_c\sim10$ eV and the injection break at $h\nu_m\sim10^{-5}$ eV. 
This interpretation is consistent with $v_{\rm flow}\sim 4000$ km s$^{-1}$, and it implies that the particle acceleration mechanism is producing a broad SED spanning at least 5 orders of magnitude.

{\em Facilities:} \facility{{\sl CXO} (ACIS, HRC)}, \facility{VLA}, \facility{GMRT}, \facility{{\sl Spitzer}}

\acknowledgements

Support for this work was provided by the National Aeronautics and Space Administration through {\sl Chandra} Award Number G03-14082 issued by the {\sl Chandra} X-ray Observatory Center, which is operated by the Smithsonian Astrophysical Observatory for and on behalf of the National Aeronautics and Space Administration under contract NAS8-03060. 
C.-Y.\ N.\ is supported by a GRF grant from the Hong Kong Government under HKU 17305416P.

\appendix

\section{A. Breaks in a Spatially-Integrated Pulsar Tail Spectrum}

A PWN tail is an outflow of relativistic electrons with a velocity $v_{\rm flow}$. 
The spectrum of synchrotron radiation from the entire observed tail (or a part of it) of a length $l_{\rm tail}$ is determined by the shape of the electron SED integrated over the coordinate $z$ along the tail ($0<z<l_{\rm tail}$) or, equivalently, over the travel time $t$ ($0<t<t_{\rm tail}$, $t_{\rm tail}=\int_0^{l_{\rm tail}} v_{\rm flow}^{-1}\,dz$).
The relativistic electrons in the outflow lose their energy to synchrotron radiation with a characteristic cooling time $t_{\rm syn} = t_{\rm cyc}\gamma^{-1}$, where $t_{\rm cyc}=9m^3c^5/(4e^4B^2) = 7.7\times 10^8 B^{-2}$ s is the cyclotron cooling time for a locally isotropic pitch angle distribution, and $\gamma$ is the Lorentz factor. 
Since high-energy particles lose their energy faster, the high-energy part of the SED steepens along the tail due to synchrotron cooling, leading to a possible break in the integrated SED at $\gamma \sim \gamma_c = t_{\rm cyc}/t_{\rm tail}\propto B^{-2} v_{\rm flow} l_{\rm tail}^{-1}$.

The shape of the integrated electron SED and the corresponding synchrotron spectrum depend on the injected SED. 
Let us assume that the SED injected at the tail's base, $z=0$, is a PL with slope $p$, an SED described by $dN_e/d\gamma \propto \gamma^{-p}$ at $\gamma_{m} < \gamma < \gamma_{M}$, and that this SED evolves along the tail due to synchrotron cooling only, without any re-acceleration.
The shape of the integrated SED depends on the relationship between $\gamma_c$ and the boundaries $\gamma_m$ and $\gamma_M$. 
If $\gamma_c > \gamma_M$, then cooling plays no role, i.e., the integrated SED $\propto \gamma^{-p}$. 
If $\gamma_m \ll \gamma_c \ll \gamma_M$ (the slow cooling regime), then the integrated SED $\propto \gamma^{-p-1}$ at $\gamma_c \ll \gamma \lesssim \gamma_M$, but it remains uncooled at $\gamma \lesssim \gamma_c$. 
Finally, if $\gamma_c \ll \gamma_m$ (the fast cooling regime), then the SED is cooled ($\propto \gamma^{-p-1}$) within the entire energy range of injected electrons, $\gamma_m \lesssim \gamma \lesssim \gamma_M$, while it becomes $\propto \gamma^{-2}$ at lower energies $\gamma_c \lesssim \gamma \lesssim \gamma_m$ due to supply of cooling electrons from higher energies. 
Thus, in both the slow and the fast cooling regimes we have two SED regions in the range min($\gamma_m, \gamma_c) < \gamma < \gamma_M$ with different slopes.
Each of these SEDs produces a synchrontron spectrum which can be approximated by a broken PL with break frequencies $\nu_m$ and $\nu_c$ ($\nu_{m,c} \approx \nu_{\rm cyc}\gamma_{m,c}^2$, where $\nu_{\rm cyc} = eB/(2\pi mc)$ is the cyclotron frequency). 
The spectral flux $F_\nu \propto \nu^{1/3}$ at the low-frequency ends of the spectra, $\nu \ll {\rm min}(\nu_m,\nu_c)$, in both cases, while the slopes of $F_\nu$ in the two other spectral regions are determined by the corresponding SED slopes.

The shapes of the synchrotron spectra for the slow and fast cooling regimes at $p>2$ were derived by Sari et al.\ (1998) who considered a similar problem for gamma-ray bursts (see also Piran 2004). 

In the slow cooling regime ($\nu_c \gg \nu_m$)
\begin{equation}
F_\nu = F_{\nu_m}\left\{
\begin{array}{lc}
        (\nu/\nu_m)^{1/3}      &       \nu < \nu_m     \\
        (\nu/\nu_m)^{-(p-1)/2} &       \nu_m < \nu < \nu_c     \\
        (\nu_c/\nu_m)^{-(p-1)/2} (\nu/\nu_c)^{-p/2}    &       \nu_c < \nu <\nu_M.    \\
\end{array}
\right.
\end{equation}
In this regime the spectrum $F_\nu$ is maximal at $\nu\approx \nu_m$, while $\nu F_\nu$ peaks at $\nu\sim \nu_c$ (for $2<p<3$).

In the fast cooling regime ($\nu_c \ll \nu_m)$ 
\begin{equation}
F_\nu = F_{\nu_c} \left\{
\begin{array}{lc}
	(\nu/\nu_c)^{1/3} 	&	\nu < \nu_c	\\
	(\nu/\nu_c)^{-1/2}	&	\nu_c < \nu < \nu_m	\\
	(\nu_m/\nu_c)^{-1/2} (\nu/\nu_m)^{-p/2}	&	\nu_m < \nu <\nu_M.	\\
\end{array}
\right.
\end{equation}
In this case, the spectral flux peaks at $\nu\approx \nu_c$, while most of the energy is emitted at $\nu\sim \nu_m$.

Although these equations are inaccurate around the break frequencies $\nu_c$ and $\nu_m$, they correctly describe the spectral slopes in broad frequency ranges if $\nu_m$, $\nu_c$ and $\nu_M$ are substantially different from each other.
We should also note that the spectral slopes can differ from the above values if additional acceleration (such as that caused by turbulent magnetic reconnection) is occurring in the emitting region (see, e.g., Xu \& Zhang 2017, Xu et al.\ in prep.).

\section{B. Magnetic Field Estimate with Allowance for Synchrotron Cooling}

As described in Appendix A, the electron SED in the slow cooling regime is
\begin{equation}
\frac{dN}{d\gamma} = K \left\{
\begin{array}{lc}
	\gamma^{-p}, & \gamma_m < \gamma < \gamma_c	\\
	\gamma_c \gamma^{-p-1}, & \gamma_c < \gamma < \gamma_M 	\\
\end{array}
\right.
\end{equation}
where $K$ is the normalization factor.
The total energy in the electrons is
\begin{multline}
W_e = m c^2 \int_{\gamma_m}^{\gamma_M} \frac{dN}{d\gamma} \gamma d\gamma \\
= K m c^2 \left[ C_{2-p}(\gamma_m,\gamma_c) + \gamma_c C_{1-p}(\gamma_c,\gamma_M) \right]\,,
\end{multline}
where 
\begin{equation}
C_q(x_1,x_2) = \int_{x_1}^{x_2} x^{q-1} dx = \left\{
\begin{array}{lc}
        (x_2^q - x_1^q)q^{-1}, & q \neq 0 \\
        \ln (x_2/x_1), & q=0. \\        
\end{array}
\right.
\end{equation}

We assume that the emission spectrum is observed in the range $\nu_1 < \nu < \nu_2$ within the cooled part of the spectrum, $\nu_m < \nu_c < \nu_1 < \nu_2 < \nu_M$.
The synchrotron luminosity in this range is 
\begin{multline}
L(\nu_1,\nu_2) = \frac{4}{3} c \sigma_T \frac{B^2}{8\pi} \int_{\gamma_1}^{\gamma_2} \frac{dN}{d\gamma}  \gamma^2 d\gamma \\
=\frac{c \sigma_T B^2 K}{6\pi} \gamma_c C_{\rm 2-p}(\gamma_1,\gamma_2)\,,
\end{multline}
where $\gamma_1$ and $\gamma_2$ are the Lorentz factors of electrons radiating at $\nu_1$ and $\nu_2$, respectively.
Substituting the normalization $K$ from Equation (B2) into Equation (B4), we obtain
\begin{equation}
L(\nu_1,\nu_2) = \frac{\sigma_T B^2}{6\pi m c} W_e \frac{ \gamma_c C_{2-p}(\gamma_1,\gamma_2) }{ C_{2-p}(\gamma_m,\gamma_c) + \gamma_c C_{1-p}(\gamma_c,\gamma_M) }.
\end{equation}
Equation (B5) allows us to express the magnetization $\sigma=W_B/W_e= B^2V/(8\pi W_e)$ (where $V$ is the volume occupied by the radiating electrons in magnetic field $B$) in terms of the luminosity:
\begin{equation}
\sigma = \frac{\sigma_T B^4 V}{48\pi^2 m c L(\nu_1,\nu_2)} \frac{ \gamma_c C_{2-p}(\gamma_1,\gamma_2) }{ C_{2-p}(\gamma_m,\gamma_c) + \gamma_c C_{1-p}(\gamma_c,\gamma_M) }\,.
\end{equation}
Taking into account that
\begin{equation}
C_q(\gamma_a,\gamma_b) = \frac{1}{2} \left( \frac{2\pi m c}{e B} \right)^{q/2} C_{q/2}(\nu_a,\nu_b)
\end{equation}
and $p=2\Gamma-2$ (where $\Gamma$ is the photon index in the cooled part of the spectrum), we obtain the following equation for the magnetic field:
\begin{equation}
B = \left[ \frac{\sigma L(\nu_1,\nu_2) }{V} \mathcal{A} \frac{ C_{2-\Gamma}(\nu_m,\nu_c) + \nu_c^{1/2} C_{1.5-\Gamma}(\nu_c,\nu_M) }{\nu_c^{1/2} C_{2-\Gamma}(\nu_1,\nu_2) } \right]^{2/7}\,,
\end{equation}
where 
\begin{equation}
\mathcal{A} = \frac{18\pi m^3 c^5}{e^4} \left( \frac{e}{2\pi m c} \right)^{1/2}.
\end{equation}

In the fast cooling regime, the electron SED can be described by 
\begin{equation}
\frac{dN}{d\gamma} = K \left\{
\begin{array}{lc}
	 \gamma^{-2}, & \gamma_c < \gamma < \gamma_m	\\
	 \gamma_m^{p-1} \gamma^{-p-1}, & \gamma_m < \gamma < \gamma_{\rm max}	\\
\end{array}
\right.
\end{equation}
assuming $\nu_c < \nu_m < \nu_1 < \nu_2<\nu_M$.  
In this case the same derivation gives
\begin{equation}
B = \left[ \frac{\sigma L(\nu_1,\nu_2) }{V}\mathcal{A} \frac{ C_0(\nu_c,\nu_m) + \nu_m^{\Gamma-1.5}C_{1.5-\Gamma}(\nu_m,\nu_M) }{ \nu_m^{\Gamma-1.5} C_{2-\Gamma}(\nu_1,\nu_2) } \right]^{2/7}.
\end{equation}

\end{document}